\documentclass[%
 reprint,
 amsmath,amssymb,
 aps,
prc
]{revtex4-2}

\usepackage{graphicx}% Include figure files
\usepackage{dcolumn}% Align table columns on decimal point
\usepackage{bm}% bold math
\usepackage[colorlinks=true, linkcolor=blue, citecolor=blue, urlcolor=blue]{hyperref}

\usepackage{braket}
\usepackage{physics}
\usepackage{siunitx}
\usepackage{booktabs}
\usepackage{orcidlink}

\begin{document}

\preprint{APS/123-QED}

\title{Impact of octupole deformation on the nuclear electromagnetic response}

\author{Manu Kanerva\orcidlink{0009-0005-5041-9277}}
 \email{manu.v.kanerva@jyu.fi}
\author{Markus Kortelainen\orcidlink{0000-0001-6244-764X}}%
 \email{markus.kortelainen@jyu.fi}
\affiliation{%
 Department of Physics, University of Jyväskylä, P.O. Box 35, FI-40014 Jyväskylä, Finland
}%

\date{\today}

\begin{abstract}
\begin{center}                          % Added this
\begin{minipage}{0.8\textwidth}       % And this to center the abstract
\begin{description}
\item[Background]
Properties of giant dipole resonances, along with other nuclear resonances, provide valuable tools for refining theoretical models as they reflect collective features of nuclear matter. Among such collective phenomena is octupole deformation, whose impact on resonance features, however, is less studied.
\item[Purpose]
Investigate the effect of reflection-symmetry-breaking octupole deformation on electric and magnetic transition strengths in atomic nuclei.
\item[Methods]
Calculations were performed using linear response theory with the iterative finite amplitude method to solve quasiparticle random phase approximation--type equations. Underlying ground-state solutions were obtained within the framework of axially symmetric Skyrme--Hartree--Fock--Bogoliubov (HFB) using three different Skyrme functionals.
\item[Results]
Electric and magnetic multipole responses were calculated for octupole-deformed even-even Rn, Ra, Th, U, Pu, and Cm isotopes. Calculations were performed on top of two distinct deformed ground-state solutions: one constrained to conserve parity, and the other allowing parity breaking. Sum rules were calculated from $M1$ transition strengths and compared with the expected correlations to certain ground-state properties.
\item[Conclusions] 
Based on our results, the octupole deformation has only a modest effect on the transition strengths in the resonances. In turn, $M1$ transition strengths have a greater effect at lower energies ($0\,\text{--}\,8\,\text{MeV}$), which encourages further investigation. Isoscalar $E3$ transition strength was confirmed to have a significant contribution from the rotational Nambu--Goldstone mode in the parity-breaking HFB solution, and thus, removing it was found necessary.
\end{description}
\end{minipage}
\end{center}
\end{abstract}

\maketitle

%\tableofcontents

\section{Introduction}
\label{sec:Introduction}
Self-consistent mean-field descriptions \cite{Ring1980, Bender2003, EDF_book2019} have been a popular tool for studying larger open-shell nuclei, which are still beyond the current reach of \textit{ab initio} methods \cite{Hergert2020}. In mean-field models, the nucleus -- a complex system of many particles that all interact with each other -- is described as a system of many particles moving independently in the mean field created by themselves. A central aspect of such descriptions is finding a solution with the lowest achievable energy, which often leads to spontaneous symmetry breaking, thereby lowering the solution's energy by incorporating additional correlations within itself \cite{Ring1980}. For instance, this can produce deformed solutions with broken rotational symmetry. Solutions with broken symmetries are referred to as intrinsic-frame solutions, in contrast to symmetry-restored laboratory-frame solutions \cite{Sheikh2021}.

Most nuclei are considered to retain axial and reflection symmetry in the intrinsic frame, exhibiting either a spherical shape for closed-shell nuclei or quadrupole deformation -- prolate or oblate -- for open-shell nuclei. However, a nucleus becomes pear-shaped if reflection symmetry is broken by axially symmetric octupole deformation. This can happen in specific regions of the nuclear chart with nucleon numbers $Z$ or $N \!\approx\! 34, 56, 88, 134$, which are just above the shell-closing magic numbers. These octupole magic numbers are associated with possible couplings between closely spaced single-particle orbitals near the Fermi surface, satisfying $\Delta j = \Delta l = 3$ \cite{Butler1996}. However, as recently studied in Ref.~\cite{Chen2021} using the Skyrme--Hartree--Fock--Bogoliubov (HFB) model, this condition alone is insufficient to determine whether the ground state is octupole-deformed.

To date, substantial experimental evidence supports permanent octupole deformation of even-even $^{224}\mathrm{Ra}$ and $^{226}\mathrm{Ra}$ isotopes \cite{Butler2020}. Additionally, a few more candidates for nuclei with permanent or vibrational octupole deformation have been identified \cite{Butler2020}. Meanwhile, global theoretical analyses using mean-field models -- such as Skyrme--HFB and relativistic Hartree--Bogoliubov (RHB) \cite{Cao2020, Agbemava2016}, and three-dimensional (3D) Skyrme--Hartree--Fock (HF) + BCS \cite{Ebata2017} for even-even nuclei, as well as the recent Skyrme parametrization BSkG3, applied within the Skyrme--HFB framework on a 3D cubic mesh for even-even, odd-A, and odd-odd nuclei \cite{Grams2023} -- predict roughly a dozen or more nuclei with either static or vibrational octupole shapes, including those already confirmed experimentally.

In the study of nuclear structure, the response of the nucleus to the electromagnetic field is a key observable. In this context, collective shape-vibrational excitations with electric multipole transitions appear at both low and high energies, associated with bound and resonance states of the nucleus \cite{Bohr1975, Ring1980}. At low energies, collective shape-vibrational excitations around the nuclear equilibrium shape may arise, standing out from single-particle excitations because of their larger reduced transition strengths. Collective high-energy excitations can likewise be described as vibrations around the equilibrium shape. In contrast to low-energy excitations, their spectrum typically shows a high density of states with a continuous character, providing a clear signature of collectivity. Because of their strong collectivity, resonance characteristics can be used to constrain and test descriptions of nuclear matter, as they correlate with its bulk properties \cite{Roca-Maza2018}.

The most extensively studied and best-known resonances are the electric isovector (IV) giant dipole resonances (GDRs) \cite{Berman1975}, which can phenomenologically be described as an out-of-phase linear vibration of neutrons against protons. Additionally, higher- and lower-order electric giant multipole resonances have also been identified and studied. The magnetic dipole response also exhibits collective excitations, including spin-flip resonances around the neutron separation energy and the low-energy scissors resonance (SR). The latter was predicted \cite{Hilton1976, Suzuki1977, Lo_Iudice_1978} and phenomenologically described as a rotational, out-of-phase vibration of neutrons against protons (scissors-like motion) and can be observed in quadrupole-deformed nuclei \cite{Bohle1984, Heyde2010}.

These resonances, especially the GDRs, play an important role in the astrophysical rapid neutron-capture process (r-process), particularly during the freeze-out, when $(n,\gamma)$--$(\gamma,n)$ equilibrium breaks down \cite{Cowan2021}. Consequently, the electric dipole ($E1$) strength function is among the key inputs in simulations predicting the abundances of heavy elements, along with other crucial parameters such as the initial electron fraction $Y_e$, which strongly influences the production and final abundances of certain long-lived actinides \cite{Eichler2019}.

Small-amplitude vibrational excitations, such as those discussed above, can be studied computationally by applying linear response theory to the mean-field solution of the nuclear ground state. Such calculations can be carried out within the framework of the random phase approximation (RPA) or its quasiparticle extension (QRPA) \cite{Ring1980, Nakatsukasa2016}, which, however, becomes computationally demanding for open-shell deformed nuclei when solved in its traditional matrix form. This difficulty can be mitigated by solving the corresponding linear response equations iteratively, for instance by using the finite amplitude method (FAM), which was developed \cite{Nakatsukasa2007, Avogadro2011} as an equivalent approach to the matrix-formulated (Q)RPA response equations. As a numerically efficient method for studying larger open-shell nuclei, FAM has been applied to a wide range of nuclear phenomena, including the nuclear response to electromagnetic fields \cite{Oishi2016, Tong2024}, collective inertia along fission paths \cite{Washiyama2021}, Thouless--Valatin inertias of Nambu--Goldstone (NG) modes \cite{Hinohara2015, Hinohara2016, Petrik2018}, and $\beta$ decays \cite{Shafer2016, Hinohara2022}.

In this work, we investigate the effects of octupole deformation on the nuclear response to the electromagnetic field. We conduct our investigation by allowing the intrinsic state of the nucleus to break rotational and reflection symmetries at the mean-field level, while conserving axial symmetry. Within these symmetry restrictions, we compare the responses of two distinct mean-field ground-state solutions: one with conserved reflection symmetry and the other with broken reflection symmetry. A study with a similar idea was recently carried out, focusing on comparing a tetrahedral shape with pure $Q_{32} \neq 0$ to spherical and pear-shaped Zr isotopes \cite{Zhao2024}.

Our study focuses on the actinide region of the nuclear chart, which contains multiple nuclei predicted to exhibit either permanent or vibrational pear shapes. More specifically, we investigate even-even Rn, Ra, Th, U, Pu, and Cm isotopes with mass numbers $A=222\,\text{--}\,230$ and $A=288\,\text{--}\,290$. Recently, a comprehensive study \cite{Tong2024} employing the FAM formalism also examined actinide nuclei, including Pu isotopes from the proton to the neutron drip line and the octupole-deformed $^{226}\text{Ra}$, but there the remaining cases considered were mostly actinides with $A\!\approx\!230\,\text{--}\,250$, which are not predicted to exhibit octupole deformation. For the nuclei studied in the present work, essentially no experimental data are currently available. Nevertheless, with upcoming developments in experimental facilities employing inverse kinematics with storage rings -- a technique already used to measure giant resonances \cite{Zamora2016} -- it may become possible to perform direct measurements of neutron capture on the corresponding short-lived nuclei (see, e.g., Refs.~\cite{Reinhart2014, Reifarth2017, Cooper2024}). Such measurements could provide data suitable for comparison with the predictions presented in this work.

To carry out this research, we use the HFBTHO code \cite{hfbtho_v2_00} to study the potential energy surfaces (PES) of the selected nuclei. Then, using the resulting ground-state solutions, we employ the FAM solver \cite{Stoitsov2011, Kortelainen2015}, which is built on top of the HFBTHO code, to study the nuclear response to electromagnetic fields. Based on the FAM calculations, we compute the transition strengths for isovector $E1$, $E2$, and $E3$, isoscalar (IS) $E2$ and $E3$, as well as $M1$ transitions, and additionally determine the photoabsorption cross sections and selected sum rules.

This article is organized as follows. Section~\ref{sec:Theory} introduces the Skyrme--HFB and FAM-QRPA formalisms used to study nuclear ground-state properties, excitations, and responses to electromagnetic fields. In Sec.~\ref{sec:Numerical_details}, we describe the numerical details and parameters employed in this work. In Sec.~\ref{sec:Results}, we present and analyze the obtained results. Finally, Sec.~\ref{sec:Conclusions} provides our conclusions. The Supplemental Material \cite{Supplemental_Material} (see also reference \cite{Scamps2013} therein) includes additional figures of the results and further less essential analyses.

\section{Theoretical framework}
\label{sec:Theory}
We study nuclei within a framework restricted to axial symmetry, allowing the breaking of reflection symmetry. We determine the ground-state properties of nuclei using the Skyrme--HFB model. Excited states and their transitions to the ground state are studied based on the obtained ground-state solutions, using FAM to solve a self-consistent set of equations corresponding to the matrix formulation of the QRPA. In the following sections, we briefly review these theoretical models as well as the treatment of the spurious NG modes.

\subsection{Skyrme--HFB model}
\label{sec:HFB+EDF}
In HFB framework \cite{Ring1980}, pairing correlations among nucleons can be taken into account, which is particularly important in deformed, open-shell nuclei. As the HFB model is based on the variational principle, it yields an HFB wave function $\ket{\Phi}$ that approximates the nuclear ground state with the lowest achievable energy $E$.

In this work, we are also considering constrained HFB solutions with specific deformations characterized by multipole moments of different orders. Since we are restricted to axially symmetric shapes, we are only considering multipole moments defined using axially symmetric spherical harmonics, $\hat{Q}_l \equiv \hat{Q}_{l,0} = r^l Y_{l, 0}(\theta, \phi)$ \cite{hfbtho_v2_00}. When these constraints are applied, a constrained HFB solution is obtained by minimizing the energy \cite{hfbtho_v2_00}
\begin{equation}
    \label{eq:HFB-constraint}
    E' = E - \sum_l \lambda_l \Bigl(\bigl\langle\hat{Q}_l \bigr\rangle - Q_l \Bigr),
\end{equation}
which represents a linear constraint formulation. In the HFBTHO, the values of Lagrange multipliers $\lambda_l$ are updated at each iteration step as described in Ref.~\cite{Younes2009}. Here, $Q_l$ is the constraint for the multipole moment of order $l$, and $\langle\hat{Q}_l \rangle$ is the expectation value of the corresponding multipole-moment operator. In this work, constraints are always imposed on the center of mass ($l=1$) with $Q_1=0$, and on the quadrupole ($l=2$) and octupole ($l=3$) moments when studying the PESs, in order to find two distinct ground-state solutions: the unconstrained ground state with $Q_3 \neq 0$ and the constrained ground state with $Q_3 = 0$.

In the Skyrme--HFB model, the density-dependent energy functional, from which the total energy is obtained via $E = \int \! d^3\textbf{r} \, \mathcal{E}(\textbf{r})$, can be separated into different contributions as \cite{Bender2003}
\begin{equation}
\label{eq:EDF_main}
    \mathcal{E} = \mathcal{E}_\text{kin} + \mathcal{E}_\text{Sk} + \mathcal{E}_\text{Coul} + \mathcal{E}_\text{pair}.
\end{equation}
Here, the first term corresponds to kinetic energy, and the remaining terms represent potential energy, consisting of Skyrme, Coulomb, and pairing contributions, respectively. Here and in the following, the dependency on densities is left implicit, although spatial dependency may be expressed when relevant.

The Skyrme term, representing strong nuclear interaction in the particle-hole channel, can be decomposed into time-even and time-odd components, $\mathcal{E}_\text{Sk} = \mathcal{E}^\text{even}_\text{Sk} + \mathcal{E}^\text{odd}_\text{Sk}$, referring to characteristics of densities from which those are constructed. Only the time-even part contributes to the ground state of even-even nuclei. However, when going beyond the mean-field level, as in time-reversal-symmetry-breaking linear-response calculations, the time-odd part also contributes and should therefore be included. Time-even and time-odd parts of the Skyrme potential can be written as \cite{Bender2003, hfbtho_v2_00}
\begin{subequations}
\label{eq:EDF_Skyrme}
    \begin{equation}
    \label{eq:EDF_Skyrme-even}
    \begin{split}
        \mathcal{E}^\text{even}_\text{Sk} = &\sum_{t} C^{\rho\rho}_t\rho^2_t
                                + C^{\rho\tau}_t \rho_t \tau_t
                                + C^{JJ}_t \textbf{J}^2_t\\
                                &+ C^{\rho\Delta\rho}_t \rho_t \Delta \rho_t
                                + C^{\rho\nabla J}_t \rho_t \nabla \cdot \textbf{J}_t,
    \end{split}
    \end{equation}
    \begin{equation}
    \label{eq:EDF_Skyrme-odd}
    \begin{split}
        \mathcal{E}^\text{odd}_\text{Sk} = &\sum_{t} C^{ss}_t \textbf{s}^{2}_t
                                +  C^{s \Delta s}_t \textbf{s}_t \cdot \Delta \textbf{s}_t
                                + C^{sT}_t \textbf{s}_t \cdot \textbf{T}_t\\
                                &+ C^{\nabla s \nabla s}_t (\nabla \cdot \textbf{s}_t)^2
                                + C^{jj}_t \textbf{j}^2_t
                                + C^{s \nabla j}_t \textbf{s}_t \cdot \nabla \cross \textbf{j}_t.
    \end{split}
    \end{equation}
\end{subequations}
Here, the $C$ coefficients are coupling constants, of which $C^{\rho \rho}_t$ and $C^{ss}_t$ are density dependent ($C^{\rho \rho}_t = C^{\rho \rho}_t[\rho_0]$ and $C^{ss}_t = C^{ss}_t[\rho_0]$). There exist many parametrizations of these coefficients, obtained by adjustment to empirical and possibly other selected data. $\rho$, $\tau$, and $\mathbf{J}$ denote matter, kinetic-energy, and spin-current densities, and $\textbf{s}$, $\textbf{j}$, and \textbf{T} represent spin, current, and spin-kinetic densities. The summation over the index $t$ accounts for isoscalar ($t=0, \rho_0 = \rho_\text{n} + \rho_\text{p}$) and isovector ($t=1, \rho_1 = \rho_\text{n} - \rho_\text{p}$) components of the densities.

The Coulomb term of the energy-density functional (EDF) (\ref{eq:EDF_main}) is split into direct and exchange contributions. The direct part is computed exactly as 
\begin{equation}
\label{eq:EDF_Coul_dir}
    \mathcal{E}^\text{dir}_\text{Coul}(\textbf{r}) = \frac{e^2}{2} 
                                    \int d^3\textbf{r}' \frac{\rho_\text{p}(\textbf{r}') \rho_\text{p}(\textbf{r})}{|\textbf{r} - \textbf{r}'|}            ,
\end{equation}
where the index p denotes protons. For more details on handling the singularity in Eq.~(\ref{eq:EDF_Coul_dir}) and its numerical implementation, see Ref.~\cite{hfbtho_v2_00}. The exchange part is obtained as the Slater approximation, which reduces the otherwise heavy computational cost of the nonlocal densities consisting exchange term.

The pairing term is calculated as
\begin{equation}
\label{eq:EDF_pairing}
    \mathcal{E}_\text{pair}(\textbf{r}) = \sum_q V^q_0 \biggl( 
    1 - \alpha\frac{\rho_0(\textbf{r})}{\rho_\text{c}} \biggr) \tilde{\rho}_q^2(\textbf{r}),
\end{equation}
where sum over $q$ accounts for neutron ($q=n$) and proton ($q=p$) pairing, and $V^q_0$ denotes the pairing strength of each type. $\rho_0(\textbf{r})$ is the local isoscalar density, $\tilde{\rho}_q^2(\textbf{r})$ is the pairing density matrix of particle type $q$, and \mbox{$\rho_\text{c} = 0.16$ $\text{fm}^{-3}$} is the saturation density of nuclear matter. The parameter $\alpha$ determines the character of the pairing interaction and can be set to any value between 0 and 1.

\subsection{FAM-QRPA method}
\label{sec:FAM}
Here, we briefly recapitulate the FAM-QRPA method for the study of nuclear excited states. For more details, see the original formulation, in which it was applied to HF+RPA \cite{Nakatsukasa2007}, its first quasiparticle extension for HFB+QRPA \cite{Avogadro2011}, and the version most relevant to this work \cite{Stoitsov2011, Kortelainen2015}.

The time-dependent HFB (TDHFB) equation
\begin{equation}
\label{eq:TDHFB}
    i \frac{\partial a_\mu(t)}{\partial t} = \bigl[\hat{H}(t) + \hat{F}(t), a_\mu(t) \bigr]
\end{equation}
can serve as the starting point of the FAM-QRPA method, from which the FAM-QRPA equations can be derived. Here, $\hat{F}(t)$ is an arbitrary one-body operator representing a weak external field, $\hat{H}(t) = \hat{H}_0 + \delta \hat{H}(t)$ is the time-dependent Hamiltonian, where the oscillation $\delta \hat{H}(t)$ around the static solution $\hat{H}_0$ is induced by the external field, and $a_\mu(t)$ is the time-dependent quasiparticle operator, which is, similarly, expressed as an oscillation around the corresponding time-independent operator.

By substituting the detailed expressions for $\hat{H}(t)$, $\hat{F}(t)$ and $a_\mu(t)$ into Eq.~(\ref{eq:TDHFB}), the FAM-QRPA equations take the form
\begin{samepage}
\begin{subequations}
\label{eq:FAM}
    \begin{equation}
    \label{eq:FAM_forward}
        (E_\mu + E_\nu - \omega) X_{\mu \nu}(\omega) + \delta H^{20}_{\mu \nu}(\omega) = F^{20}_{\mu \nu}(\omega),
    \end{equation}
    \vspace{-20pt}
    \begin{equation}
    \label{eq:FAM_backward}
        (E_\mu + E_\nu + \omega) Y_{\mu \nu}(\omega) + \delta H^{02}_{\mu \nu}(\omega) = F^{02}_{\mu \nu}(\omega),
    \end{equation}
\end{subequations}
\end{samepage}
which form a set of nonlinear equations since $\delta H^{20}$ and $\delta H^{02}$ depend on $X$ and $Y$. Therefore, the FAM-QRPA equations (\ref{eq:FAM}) are solved iteratively, with the amplitudes $X$ and $Y$ updated at each iteration. All other quantities in Eq.~(\ref{eq:FAM}), except $\delta H^{20}$ and $\delta H^{02}$, can be computed from the static HFB solution and the definition of the external field. For further details on solving FAM-QRPA equations in this work, see Ref.~\cite{Kortelainen2015}.

\subsection{Transition strengths and sum rules}
\label{sec:Strength_function}
FAM-QRPA equations (\ref{eq:FAM}) are solved for a given energy $\omega$, for which the complex value is typically introduced as $\omega \rightarrow \omega_\gamma = \omega + i \gamma$, where $\gamma$ corresponds to Lorentzian smearing with $\Gamma \!\approx\! 2\gamma$. The strength function is calculated from the solution as
\begin{equation}
\label{eq:S(F;omega)}
\begin{split}
    S(\hat{F}; \omega) &= \frac{1}{2} \sum_{\mu\nu} \bigl\{  F^{20*}_{\mu\nu} X_{\mu \nu}(\omega) + F^{02*}_{\mu\nu} Y_{\mu \nu}(\omega) \bigr\}\\
                       &= \text{Tr} \bigl[f^\dagger \rho_f \bigr],
\end{split}
\end{equation}
where, in the second line, the particle-hole type field operator is assumed \cite{Avogadro2011}. Transition strength is then obtained from the strength function as
\begin{equation}
\label{eq:TrnStr}
\begin{split}
    \frac{dB(\omega; \hat{F})}{d\omega} &= \sum_{n>0} \bigl|\bra{\Phi_n} \hat{F}\ket{\Phi_0}\bigr|^2 \delta(\omega - E_n)\\
                                        &= -\frac{1}{\pi} \text{Im}\bigl[ S(F; \omega) \bigr],
\end{split}
\end{equation}
where in the first line of the expression, $n$ refers to excited states $\ket{\Phi_n}$ with energy $E_n$, and $\ket{\Phi_0}$ is the HFB ground state with the assumed ground-state energy of zero.

To calculate transition strengths for electric and magnetic transitions with a given multipole $L$ and its $z$-projection $K$, the appropriate definitions for $\hat{F}$ must be used. For these operators, the field takes the particle-hole form $\hat{F} = \sum_{kl} f_{kl} c^\dagger_k c_l$. In the case of electric isoscalar (IS) and isovector (IV) transitions, the matrix elements $f_{kl}$ are expressed in terms of spherical harmonics as follows
\begin{subequations}
\label{eq:EL_operators}
    \begin{equation}
    \label{eq:EL_IS}
    f^\text{IS}_{LK} = e_\text{IS} \sum_{i=1}^A r^L_i Y_{LK}(\hat{\textbf{r}}_i),
    \end{equation}
    \vspace{-20pt}
    \begin{equation}
    \label{eq:EL_IV}
    f^\text{IV}_{LK} = \sum_{i=1}^A e_{\text{IV}, \tau_i} \tau_i r^L_i Y_{LK}(\hat{\textbf{r}}_i).
    \end{equation}
\end{subequations}
Here $\tau_i = \pm 1$ for neutrons and protons, and $e_\text{IS}=eZ/A$ denotes the effective isoscalar charge, while the isovector neutron and proton effective charges are given by $e_\text{IV,n}= eZ/A$ and $e_\text{IV,p} = eN/A$.

The operator for magnetic dipole transitions is \cite{Ring1980}
\begin{equation}
\label{eq:M1}
    f^\text{M}_{1K} = \mu_N \sqrt{\frac{3}{4\pi}} \sum_{i=1}^A \biggl( g^i_s \textbf{s}^i_K +  g_l^i  \bm{l}^i_K \biggr),
\end{equation}
where $\mu_N$ is the nuclear magneton, $\textbf{s}_K$ and $\bm{l}_K$ are the $K$th components of the spin and orbital angular momentum vectors, and $g_s$ and $g_l$ are the corresponding bare gyromagnetic factors, which in this work are taken as $g_s^\text{n(p)} = -3.826 (5.586)$ and $g_l^\text{n(p)} = 0.0 (1.0)$.

The photoabsorption cross section generally includes contributions from all transition types discussed above and possibly others. However, since the electric dipole transitions are known to dominate the total cross section, we evaluate it at a given energy by taking into account only these transitions. According to Ref.~\cite{Ring1980}, the cross section is given by
\begin{equation}
\label{eq:cross-section}
    \sigma_\text{abs}(\omega) = \sum_{K=0,\pm1} \frac{4 \alpha \pi^2}{e^2}\omega \frac{dB\bigl(\omega; \hat{F}^\text{IV}_{E1} \bigr)}{d\omega}, 
\end{equation}
where $\alpha = e^2 / \hbar c$ denotes the fine structure constant.

The energy-weighted sum rule of order $k$ can be approximated as
\begin{equation}
\label{eq:m_k}
    m_k = \sum_{K=-L}^L \int_{\omega_i}^{\omega_f} \omega^k \frac{dB\bigl(\omega; \hat{F}_{LK} \bigr)}{d\omega} d\omega
\end{equation}
for any of the transition operators $\hat{F}$ defined above. Ideally, the integration limits in Eq.~(\ref{eq:m_k}) would extend from zero to $\infty$, but due to numerical limitations, practical finite limits must be used.

\subsection{Removal of Nambu--Goldstone modes}
\label{sec:NG-modes}
Nambu--Goldstone (NG) modes can arise in the linear response calculations as a symmetry-restoring zero-energy mode related to broken symmetries in the underlying static mean-field solution \cite{Ring1980}. Modes with imaginary QRPA energies $\Omega$ may also arise if the static mean-field solution is not a minimum, but instead a saddle point of the PES with respect to a relevant variable. Consequently, the calculated strength function (\ref{eq:S(F;omega)}) can be decomposed into three contributions
\begin{equation}
    S^\text{calc} = S^\text{phys} + S^\text{NG} + S^{\text{Im} \Omega},
\end{equation}
corresponding to the physical, NG, and imaginary QRPA pole contributions, which are associated with real, zero, and imaginary energies, respectively. However, due to the use of a finite basis size, the energies are not exactly so. For the remainder of this section, we disregard $S^{\text{Im} \Omega}$ and instead provide a brief recapitulation of how to remove the NG contribution from the calculated strength function within the FAM formulation.

To distinguish NG modes from physical excitations in linear response calculations, one needs to identify the broken symmetries and choose operators $\hat{\mathcal{P}}$, which are generators of the group of the recognized broken symmetries \cite{Kortelainen2020}. A conjugate operator $\hat{\mathcal{Q}}$ satisfying canonicity condition $[\hat{\mathcal{Q}}, \hat{\mathcal{P}}] = i$ is also required to remove the NG mode as it can be viewed to split into spurious and boost contributions related to these two operators.

The broken symmetries with corresponding NG modes that we consider are those related to translational invariance and rotational symmetry. Assuming axial and reflection symmetries, the NG modes may arise for operators $\hat{F}$ with $K^\pi = 0^-, 1^-$ related to broken translational invariance, or for $K^\pi = 1^+$ related to broken rotational symmetry. For translational invariance, the relevant operator is the momentum operator $\hat{P}_\text{c.m.}$, whose conjugate operator is the coordinate operator $\hat{Q}_\text{c.m.}$. For rotational symmetry, the relevant operator is the total angular momentum operator for which the suitable component must be selected based on the symmetries; in this case, it is $\hat{J}_y$ because of the applied simplex-$y$ symmetry. As is often the case, its conjugate operator is not known in advance. However, an expression for obtaining the conjugate operator within linear response theory was derived in Ref.~\cite{Hinohara2015} as
\begin{equation}
    \hat{\mathcal{Q}}_\text{NG} = -i \frac{X(0) + Y^*(0)}{2 S(\hat{\mathcal{P}}_\text{NG}; 0)}.
\end{equation}
Hence, performing the full FAM-QRPA calculation using the relevant operator ($\hat{\mathcal{P}} \rightarrow \hat{J}_y$) as an external field with zero energy ($\omega = 0$) is required for the construction of the conjugate operator. As a result of this part of the process, also, the rotational Thouless--Valatin moment of inertia along the $y$ axis is obtained as \cite{Petrik2018}
\begin{equation}
\label{eq:TV_MoI}
    M_\text{TV}^{J_y} = - S(\hat{J}_y; 0).
\end{equation}

The method for separating the NG contribution from the calculated strength function in the FAM formalism was proposed already in the first introduction to FAM in Ref.~\cite{Nakatsukasa2007}. In the current implementation \cite{Kortelainen2020}, the NG mode is removed from $X$ and $Y$ amplitudes used to calculate the strength function. Thus, physical $X$ and $Y$ amplitudes are obtained as
\begin{subequations}
    \begin{equation}
        X^\text{phys}_{\mu\nu}(\omega) = X^\text{calc}_{\mu\nu}(\omega) - \lambda_P P^{20} - \lambda_Q Q^{20},
    \end{equation}
    \vspace{-20pt}
    \begin{equation}
        Y^\text{phys}_{\mu\nu}(\omega) = Y^\text{calc}_{\mu\nu}(\omega) - \lambda_P P^{02} - \lambda_Q Q^{02},
    \end{equation}
\end{subequations}
where $\lambda_P$ and $\lambda_Q$ are determined by the canonicity condition and the requirement that physical and NG modes are orthogonal. Thus,
\begin{subequations}
    \begin{equation}
        \lambda_P = \mathcal{N}^{-1} \Bigl( \bra{Q^{20}} \ket{X^\text{calc}(\omega)} + \bra{Q^{02}} \ket{Y^\text{calc}(\omega)} \Bigr),
    \end{equation}
    \vspace{-20pt}
    \begin{equation}
        \lambda_Q = -\mathcal{N}^{-1} \Bigl( \bra{P^{20}} \ket{X^\text{calc}(\omega)} + \bra{P^{02}} \ket{Y^\text{calc}(\omega)} \Bigr),
    \end{equation}
\end{subequations}
where $\mathcal{N} = \bra{Q^{20}, Q^{02}}\ket{P^{20}, P^{02}}$ is the normalization constant, typically equal to $i$, though some deviation may occur due to the finite basis size.

\section{Numerical details}
\label{sec:Numerical_details}
\begin{table}[t]
\caption{Neutron and proton pairing strengths $V^n_0$ and $V^p_0$ for the Skyrme parametrizations used.}
\label{tab:pairing_strengths}
\begin{ruledtabular}
\begin{tabular}{lccc}
    & SkM* (MeV) & SLy4 (MeV) & UNEDF1 (MeV) \\
%\midrule
\hline
$V^n_0$ & -265.2500 & -325.2500 & -223.2780 \\
$V^p_0$ & -340.0625 & -340.0625 & -247.8960 \\
%\bottomrule
\end{tabular}
\end{ruledtabular}
\end{table}
We performed HFB+FAM-QRPA calculations using SkM* \cite{Bartel1982}, SLy4 \cite{Chabanat1998} and UNEDF1 \cite{Kortelainen2012} Skyrme functionals. The ground-state solutions of the studied nuclei were determined using the HFBTHO program \cite{hfbtho_v2_00}. The number of shells defining the basis size of harmonic oscillators was set to $N_\text{sh} = 20$ for SkM* and SLy4, and to $16$ for UNEDF1. The smaller value for UNEDF1 was used to reduce the computation time, as FAM-QRPA calculations with this functional required more iterations, and also because it matches the value for the UNEDF1 variant employed in this work and described below. In all calculations, we set the pairing cutoff to $\SI{60}{MeV}$, and used the value $\alpha = 1/2$ in Eq.~(\ref{eq:EDF_pairing}), corresponding to mixed pairing in contrast to volume and surface pairings. The neutron and proton pairing strengths used for each parametrization are given in Tab.~\ref{tab:pairing_strengths}. For the SkM* functional, the pairing strengths correspond to the values used in Ref.~\cite{Schunck2014}, which were adjusted to reproduce the $^{240}\text{Pu}$ pairing gaps, while for SLy4 they correspond to the values used in Ref.~\cite{Chen2021}. These values are also provided as the default pairing strengths for these functionals in the HFBTHO program. The pairing strengths used for \mbox{UNEDF1} were increased by 20\, \% compared to the original values of that functional, following Ref.~\cite{Bonnard2023}, to compensate for the absence of Lipkin-Nogami particle number projection (PNP), which enhances pairing correlations and was employed in the original UNEDF1 fitting.

The time-odd coupling constants used in Eq.~(\ref{eq:EDF_Skyrme-odd}) follow their original parametrizations for the SkM* and SLy4 functionals. Since UNEDF1 is formulated as a more general EDF and its time-odd coupling constants were not defined in the original fit, in this work they are obtained from the Landau parameters of the Landau--Migdal interaction; see, e.g., Ref.~\cite{Bender2002} for details. We used the values $g_0=0.4$, $g_0'=1.7$ based on the fitted values for UNEDF1 in Ref.~\cite{Sassarini_2022}.

The FAM-QRPA equations were solved, and the transition strength functions were obtained using the FAM solver \cite{Kortelainen2015}, which is built on top of the HFBTHO program. A value of $\gamma=\SI{0.5}{MeV}$ for the imaginary part of the energy was used in all calculations unless stated otherwise. To obtain the transition strengths as a function of excitation energy, the FAM-QRPA equations were typically solved at $\SI{0.3}{MeV}$ intervals from $0$ to $\SI{38}{MeV}$. However, for UNEDF1, the calculations were generally performed only up to $\SI{20}{MeV}$ because, above this energy, computations became slow or failed to converge. Additionally, a smaller step size was adopted when using a smaller $\gamma$. For $\gamma = \SI{0.5}{MeV}$, some of the $E3$ isovector calculations were performed with a step size of $\SI{0.35}{MeV}$.

\section{Results}
\label{sec:Results}
\begin{table*}
\caption{Intrinsic quadrupole moments $Q^\prime_2$ for reflection-symmetric energy minima, and quadrupole ($Q_2$) and octupole ($Q_3$) moments for reflection-symmetry-breaking HFB ground states. Deformation energies are given for spherically symmetric $E_{\text{def}}(0, 0)$ and octupole-deformed $E_{\text{def}}(Q_2, Q_3)$ solutions, relative to the total energy of the reflection-symmetric lowest-energy solution $E(Q'_2, 0)$.}
\label{tab:GS-properties}
\begin{ruledtabular}
\begin{tabular}{cccccccc}
 Nucleus & Force & $Q^\prime_2$ (b) & $Q_2$ (b) & $Q_3$ ($\mathrm{b}^{3/2}$) &
 $E(Q'_2, 0)$ (MeV) & $E_{\text{def}}(0, 0)$ (MeV) & $E_{\text{def}}(Q_2, Q_3)$ (MeV) \\
\hline\\[-9pt]
$^{222}\text{Rn}$ & SkM* & 12.509 & 12.669  & 2.830
 & -1703.679 & 0.595
& -0.438  \\
 & SLy4 & 12.112 & 12.427 & 1.127
 & -1699.971& 1.197 & -0.027  \\
$^{224}\text{Rn}$ & SkM* & 15.075 & 15.082  & 1.792
 & -1714.463 & 1.690
& -0.053  \\
[5pt]
$^{222}\text{Ra}$ & SkM* & 12.615 & 13.780  & 3.311
 & -1701.435 & 0.807
& -1.391  \\
 & SLy4 & 12.363 & 13.920 & 2.484
 & -1701.029& 1.301 & -0.700  \\
 & UNEDF1 & 12.773 & 14.133 & 2.435
 & -1707.339& 1.329 & -0.611  \\
$^{224}\text{Ra}$ & SkM* & 15.922 & 15.953  & 3.395
 & -1713.749 & 2.351
& -0.919  \\
 & SLy4 & 14.809 & 15.788 & 2.707
 & -1711.932& 2.599 & -0.659  \\
 & UNEDF1 & 15.245 & 15.972 & 2.556
 & -1718.647& 2.417 & -0.511  \\
$^{226}\text{Ra}$ & SkM* & 18.343 & 18.121  & 2.988
 & -1725.840 & 3.946
& -0.322  \\
 & SLy4 & 17.093 & 17.546 & 2.644
 & -1722.532& 3.881 & -0.423  \\
 & UNEDF1 & 17.476 & 17.683 & 2.262
 & -1729.661& 3.504 & -0.266  \\
$^{228}\text{Ra}$ & SkM* & 20.257 & 20.282  & 1.409
 & -1737.544 & 5.282
& -0.026  \\
 & SLy4 & 19.107 & 19.162 & 2.044
 & -1732.786& 5.082 & -0.129  \\
 & UNEDF1 & 19.399 & 19.378 & 1.617
 & -1740.286& 4.484 & -0.070  \\
[5pt]
$^{222}\text{Th}$ & SkM* & 10.521 & 11.125  & 3.106
 & -1695.131 & 0.218
& -1.619  \\
 & SLy4 & 10.879 & 12.626 & 2.398
 & -1697.522& 0.527 & -0.717  \\
 & UNEDF1 & 11.292 & 12.888 & 2.453
 & -1703.599& 0.563 & -0.821  \\
$^{224}\text{Th}$ & SkM* & 15.239 & 15.242  & 3.559
 & -1708.708 & 1.944
& -1.522  \\
 & SLy4 & 14.393 & 15.174 & 2.839
 & -1709.960& 2.138 & -0.943  \\
 & UNEDF1 & 14.934 & 15.387 & 2.872
 & -1716.354& 1.907 & -0.979  \\
$^{226}\text{Th}$ & SkM* & 18.802 & 17.734  & 3.620
 & -1722.446 & 4.117
& -0.783  \\
 & SLy4 & 17.635 & 17.376 & 3.115
 & -1722.243& 3.690 & -0.757  \\
 & UNEDF1 & 18.340 & 17.566 & 3.084
 & -1728.973& 3.428 & -0.714  \\
$^{228}\text{Th}$ & SkM* & 21.004 & 20.406  & 3.090
 & -1735.774 & 6.036
& -0.163  \\
 & SLy4 & 20.245 & 19.578 & 3.113
 & -1734.235& 5.437 & -0.348  \\
 & UNEDF1 & 20.783 & 19.797 & 2.962
 & -1741.264& 4.739 & -0.266  \\
[5pt]
$^{224}\text{U}$ & SkM* & 12.347 & 11.741  & 3.271
 & -1699.625 & 0.811
& -1.648  \\
 & SLy4 & 12.173 & 13.171 & 2.596
 & -1703.603& 0.735 & -0.773  \\
 & UNEDF1 & 12.831 & 13.504 & 2.752
 & -1709.794& 0.786 & -1.009  \\
$^{226}\text{U}$ & SkM* & 17.556 & 16.236  & 3.692
 & -1714.514 & 2.969
& -1.256  \\
 & SLy4 & 16.423 & 16.185 & 3.059
 & -1717.310& 2.687 & -0.853  \\
 & UNEDF1 & 17.570 & 16.504 & 3.185
 & -1723.823& 2.427 & -0.989  \\
$^{228}\text{U}$ & SkM* & 21.106 & 19.360  & 3.678
 & -1729.555 & 5.555
& -0.366  \\
 & SLy4 & 20.363 & 18.854 & 3.345
 & -1730.966& 4.905 & -0.502  \\
 & UNEDF1 & 21.163 & 19.103 & 3.415
 & -1737.786& 4.377 & -0.528  \\
[5pt]
$^{226}\text{Pu}$ & SkM* & 13.774 & 11.869  & 3.402
 & -1702.691 & 1.161
& -1.519  \\
 & SLy4 & 13.336 & 13.369 & 2.692
 & -1708.162& 0.944 & -0.659  \\
 & UNEDF1 & 14.632 & 13.510 & 2.950
 & -1714.458& 1.036 & -0.960  \\
$^{228}\text{Pu}$ & SkM* & 19.776 & 16.988  & 3.725
 & -1718.772 & 3.698
& -0.834  \\
 & SLy4 & 18.371 & 17.025 & 3.076
 & -1723.076& 3.198 & -0.539  \\
 & UNEDF1 & 20.685 & 17.291 & 3.302
 & -1729.784& 3.118 & -0.601  \\
$^{230}\text{Pu}$ & SkM* & 22.713 & 21.176  & 3.371
 & -1734.851 & 6.607
& -0.043  \\
 & SLy4 & 22.001 & 20.666 & 3.099
 & -1737.920& 5.705 & -0.165  \\
 & UNEDF1 & 23.101 & 21.265 & 3.228
 & -1744.877& 5.359 & -0.134  \\
$^{288}\text{Pu}$ & SkM* & 20.266 & 15.931  & 5.402
 & -1992.829 & 0.190
& -1.438  \\
 & SLy4 & 17.218 & 18.117 & 4.589
 & -1952.142& 0.005 & -0.849  \\
$^{290}\text{Pu}$ & SkM* & 28.325 & 22.099  & 5.798
 & -1997.272 & 1.956
& -0.509  \\
 & SLy4 & 25.124 & 22.511 & 5.050
 & -1954.144& 1.488 & -0.569  \\
 & UNEDF1 & 23.298 & 20.739 & 4.605
 & -1980.658& 0.249 & -0.796  \\
[5pt]
$^{228}\text{Cm}$ & SkM* & 14.887 & 11.699  & 3.513
 & -1704.331 & 1.456
& -1.326  \\
 & SLy4 & 14.345 & 13.373 & 2.708
 & -1711.228& 1.267 & -0.457  \\
 & UNEDF1 & 20.438 & 13.273 & 3.138
 & -1717.665& 1.424 & -0.736  \\
$^{290}\text{Cm}$ & SkM* & 32.332 & 13.996  & 5.551
 & -2015.804 & 0.860
& -1.316  \\
 & SLy4 & 19.380 & 17.595 & 4.897
 & -1978.145& 0.450 & -0.910  \\
 & UNEDF1 & 16.680 & 15.428 & 4.605
 & -2002.682& 0.224 & -1.240  \\
\end{tabular}
\end{ruledtabular}
\end{table*}
\begin{figure*}[t]
\includegraphics[width=\textwidth]{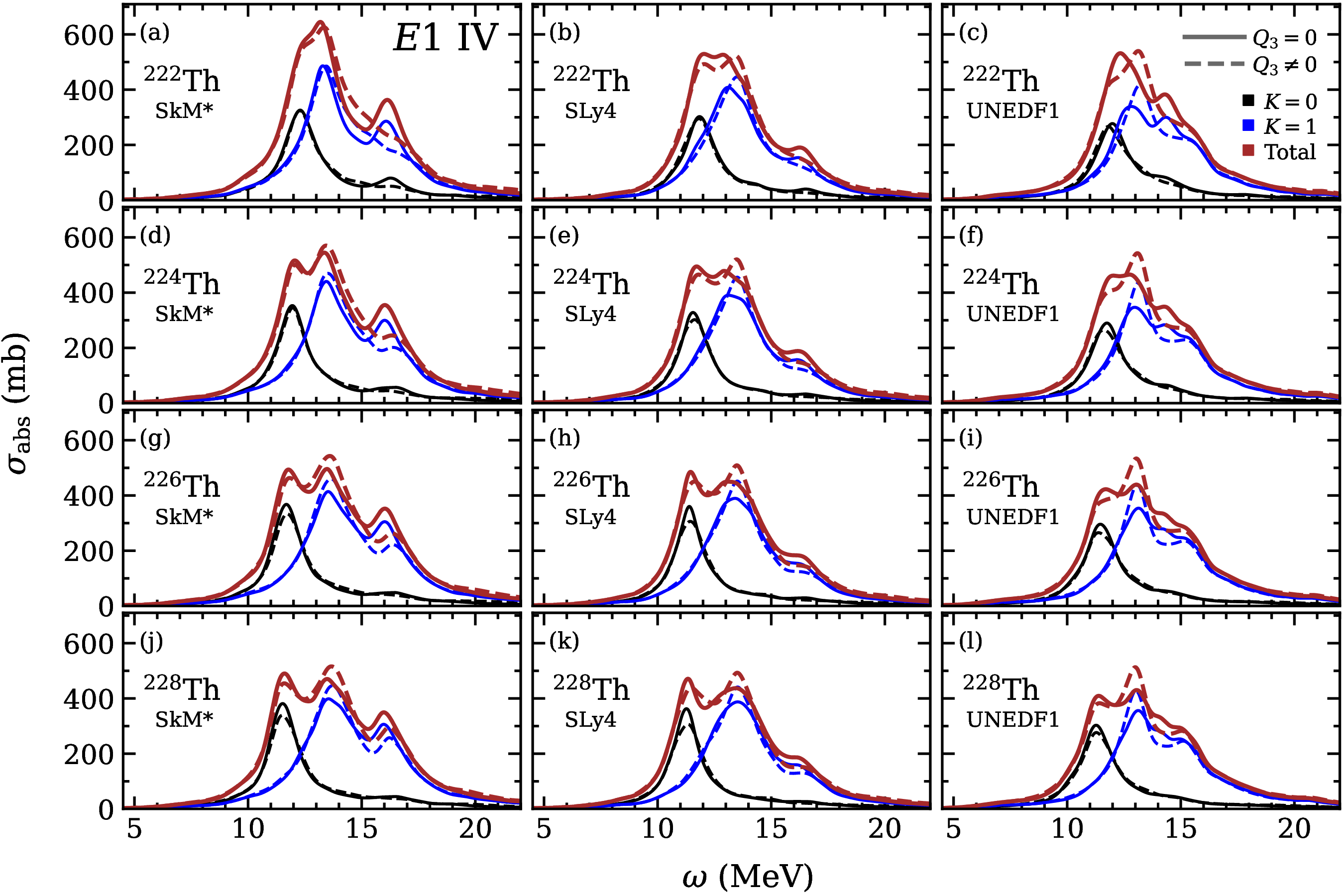}
\caption{\label{fig:E1_Th}Photoabsorption cross sections calculated from isovector (IV) electric dipole ($E1$) transition strengths for studied Th isotopes. Panels (a)--(l) show results for isotopes with mass numbers $A = 222\,\text{--}\,228$, calculated using three different Skyrme energy-density functionals, SkM*, SLy4, and UNEDF1, as indicated in each panel. Solid and dashed lines correspond to reflection-symmetry-conserving ($Q_3=0$) and reflection-symmetry-breaking ($Q_3\neq0$) HFB ground-state solutions, respectively. The $K=1$ mode is shown as the sum of the identical $K=\pm1$ modes.}
\end{figure*}
The nuclear electromagnetic response was studied on top of two distinct deformed ground-state solutions -- one with conserved and another with broken reflection symmetry. Table~\ref{tab:GS-properties} presents intrinsic multipole moments along with the corresponding total and deformation energies of these solutions. The response of these HFB ground-state solutions to electric and magnetic fields of various multipoles is discussed in the following sections, primarily by presenting typical examples of transition strength functions obtained for the studied nuclei, with the remaining results provided in the Supplemental Material \cite{Supplemental_Material}.

\subsection{\textit{E}1 transitions}
\label{sec:E1-transitions}
Photoabsorption cross sections for $E1$ transitions in all studied Th isotopes, calculated using the three Skyrme functionals employed in this research, are shown in Fig.~\ref{fig:E1_Th}. The total cross sections are spread due to quadrupole deformation, which causes a split in $K=0$ and $K=1$ modes. The $K=1$ mode lies higher in energy as expected for positive quadrupole deformations ($Q_2>0$). As the quadrupole deformation increases, the split of the total photoabsorption cross section becomes more pronounced, and in strongly enough deformed isotopes, a two-peak structure may emerge. A third peak, or so-called right shoulder, may appear around $\omega = \SI{16}{MeV}$, particularly in calculations using the SkM* functional. Previous studies \cite{Nesterenko2007} attribute this feature to the small effective isovector mass of the functional and the presence of the high-angular-momentum intruder states.

Our results in Fig.~\ref{fig:E1_Th} and in the Supplemental Material \cite{Supplemental_Material} indicate that octupole deformation has only a modest impact on the giant dipole resonances. However, some differences do emerge relatively consistently across the studied nuclei and Skyrme functionals. Specifically, the first peak in cross sections, related to the $K=0$ mode, is slightly reduced in most of the octupole deformed HFB solutions, while the second peak related to the $K=1$ mode shows an enhancement. In the SkM* results, the third peak is in some cases completely suppressed in the octupole-deformed HFB solution. However, since this is not observed in all cases, the suppression may not be caused by the octupole deformation itself, but rather as a consequence of secondary modifications it induces.

\subsection{\textit{M}1 transitions}
\label{sec:M1-transitions}
\begin{figure}[tbp]
\includegraphics[width=8.5cm]{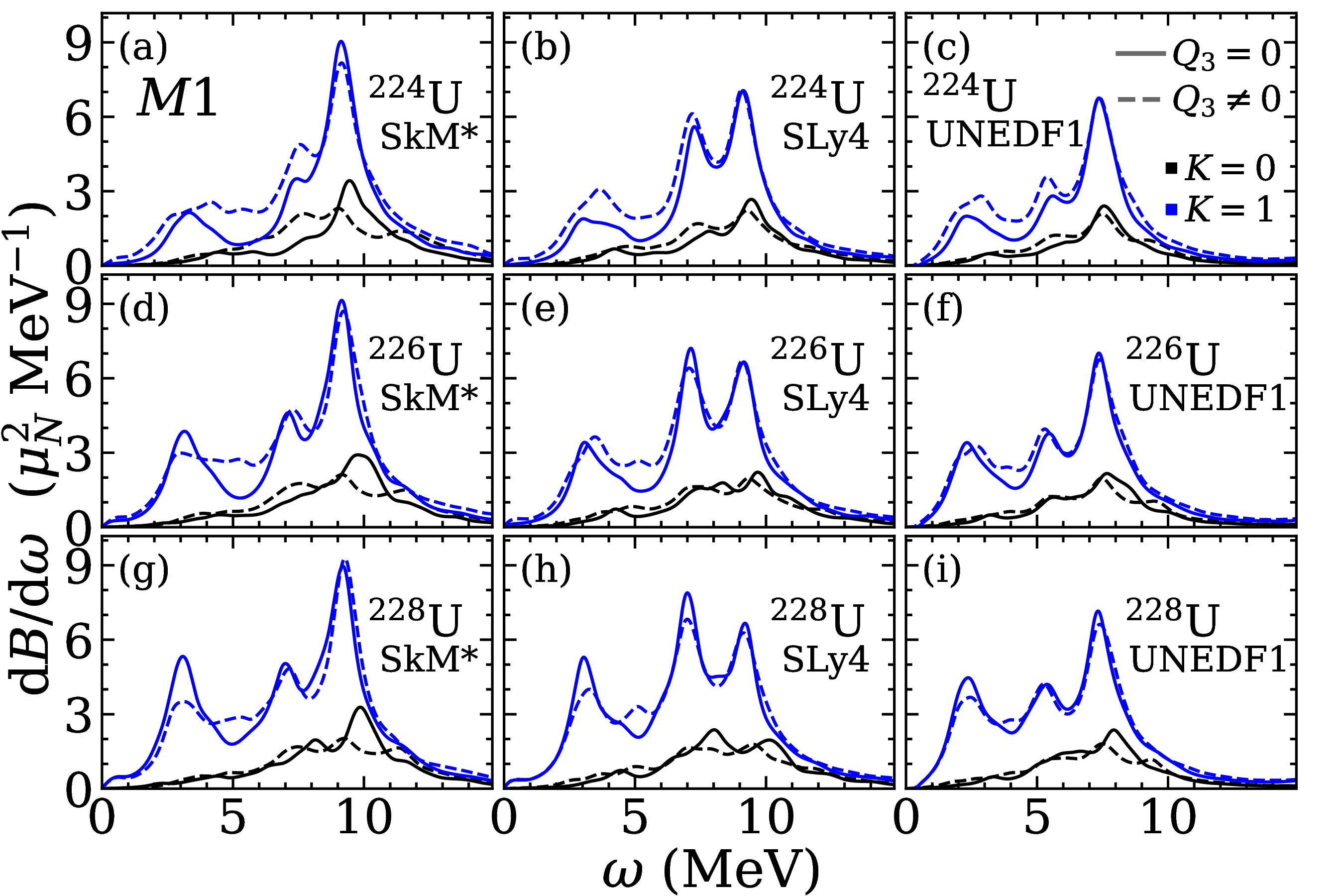}
\caption{\label{fig:M1_U}Magnetic dipole ($M1$) transition strength functions for studied U isotopes. Panels (a)--(i) show results for isotopes with mass numbers $A = 224\,\text{--}\,228$, calculated using three different Skyrme energy-density functionals, SkM*, SLy4, and UNEDF1, as indicated in each panel. Solid and dashed lines correspond to reflection-symmetry-conserving ($Q_3=0$) and reflection-symmetry-breaking ($Q_3\neq0$) HFB ground-state solutions, respectively. Modes with $K>0$ are shown as the sum of the corresponding identical positive and negative $K$ modes.}
\end{figure}
Figure~\ref{fig:M1_U} shows the $M1$ transition strength functions in U isotopes. In the UNEDF1 results, the transition strength is concentrated below $\omega = \SI{10}{MeV}$, whereas in the SkM* and SLy4, they extend slightly beyond this energy. Within these energy limits, $M1$ transitions can be grouped into two distinct regions: (i) magnetic giant dipole resonances, consisting mostly of spin-flip transitions in the range of approximately $5\,\text{--}\,\SI{10}{MeV}$, and (ii) lower-energy transitions, where a significant orbital component also appears.

In our results, the $M1$ giant resonances exhibit a two-peak structure for the $K=1$ mode with all three Skyrme functionals. This is consistent with experimental observations of two humps in many deformed nuclei \cite{Heyde2010, Richter1995}. In our results, the lower-energy peak is attributed to proton spin-flips, and the higher-energy peak to neutron spin-flips. However, earlier separable RPA (SRPA)-based studies with Skyrme EDFs \cite{Vesely2009} have found either one or two peaks in both spherical and deformed nuclei. Therefore, the two-peak structure reproduced in our results cannot be taken as evidence that the experimental trend is successfully reproduced. The $K=0$ mode, which in these deformed nuclei can be regarded as a background $M1$ mode because its transition strength is much smaller than that of the $K=1$ mode, does not exhibit as clear or consistent a pattern. Nevertheless, it displays a similar two-peak structure associated with neutron and proton transition strength distributions, following the same peak order as the $K=1$ mode, but at slightly higher energies.

In the $M1$ transition strength functions, all Skyrme functionals and both deformed HFB ground-state solutions exhibit similar overall transition strength distributions, although notable differences also emerge. For excitation energies below approximately $\SI{8}{MeV}$, the octupole-deformed HFB solution yields larger transition strengths for both the $K=0$ and $K=1$ modes. In turn, above $\SI{8}{MeV}$, the transition strengths are mostly comparable between the two deformed ground-state solutions. An exception is the SkM* functional, which exhibits, in particular, larger $K=0$ transition strengths around $\SI{10}{MeV}$ for the reflection-symmetric HFB ground state.

Around $\omega = \SI{3}{MeV}$, within the energy range of the expected scissors resonance (SR), significant transition strengths appear for both deformed ground-state solutions in the $K=1$ results. The transition strength functions of reflection-symmetric HFB solutions more often exhibit a peak-like structure around $\SI{3}{MeV}$, whereas in the octupole-deformed HFB solution, the peak is generally less pronounced (see also the Supplemental Material \cite{Supplemental_Material}), and instead resembles a plateau between $2$ and $\SI{5}{MeV}$. The scissors resonance peak is known to be bigger with larger quadrupole deformation \cite{Ziegler1990, Heyde2010}. Consistent with this, Fig.~\ref{fig:M1_U} and the Supplemental Material \cite{Supplemental_Material} show that the corresponding peak becomes more pronounced in FAM-QRPA calculations based on the reflection-symmetric HFB solution along isotopic chains, where quadrupole deformation grows with increasing neutron number. For octupole-deformed solutions, the resulting plateau remains closer to the same transition strength, regardless of the quadrupole moment along isotopic chains.

\begin{figure}[tbp]
\includegraphics[width=8.5cm]{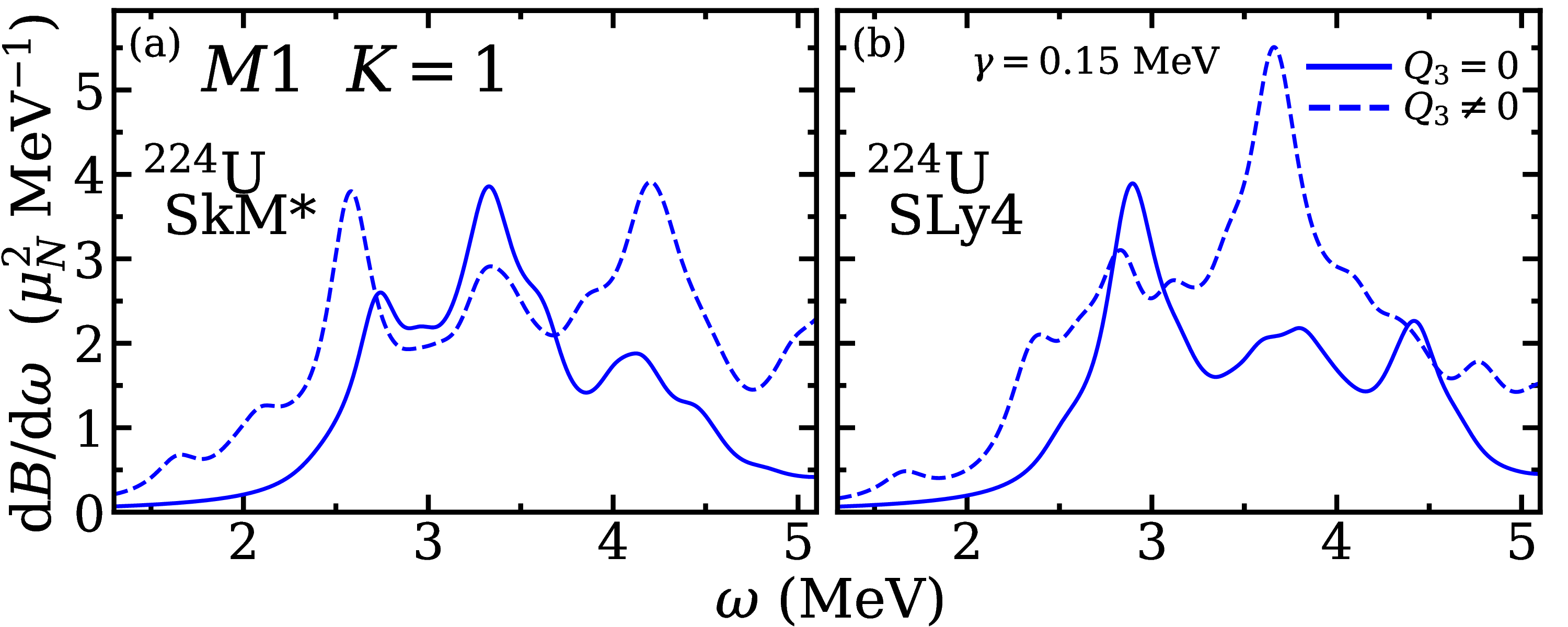}
\caption{\label{fig:M1_U224_small_gamma}Same as Fig.~\ref{fig:M1_U}, but for the $K=1$ mode of the $^{224}\text{U}$ isotope only, calculated with a smaller imaginary part of the energy ($\gamma$) using two Skyrme EDFs (SkM* and SLy4), as indicated in each panel.}
\end{figure}
To gain deeper insight into the differences at low energies, we repeated the calculations for selected cases in this energy range using a smaller imaginary part of the energy, $\gamma$. This choice corresponds to a smaller Lorentzian smearing, $\Gamma$, and thus provides a more detailed view of the underlying low-energy excitation modes. Figure~\ref{fig:M1_U224_small_gamma} shows the $K=1$ transition strength functions of $^{224}\text{U}$, calculated with reduced $\gamma$ using SkM* and SLy4 functionals. For SkM*, the reflection-symmetric HFB solution exhibits a clear peak that may correspond to the scissors resonance around $\omega = \SI{3.5}{MeV}$, whereas the pear-shaped solution shows a reduced peak. Instead, its transition strength is spread across the possible SR region, forming two distinct peaks at about $\SI{2.5}{MeV}$ and just above $\SI{4}{MeV}$. With the SLy4 functional, the fragmentation differs somewhat: the reflection-symmetric HFB solution shows a larger peak at slightly lower energy, just below $\SI{3}{MeV}$, while the pear-shaped solution exhibits a peak at higher energy, around $\SI{3.7}{MeV}$.

\begin{figure}[tbp]
\includegraphics[width=8.5cm]{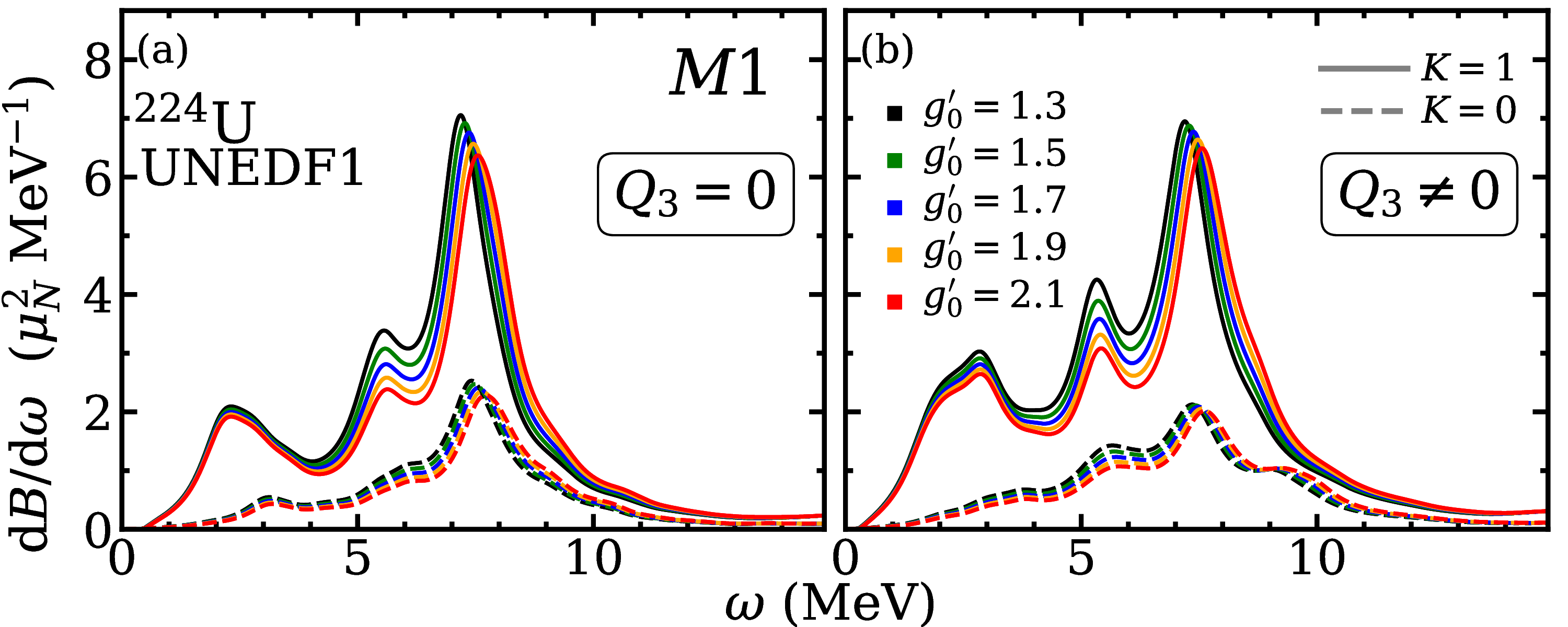}
\caption{\label{fig:M1_Landau}Similar to Fig.~\ref{fig:M1_U}, but shown only for $^{224}\text{U}$. Panel (a) displays results for the reflection-symmetric HFB solution \mbox{($Q_3=0$)}, while panel (b) corresponds to the reflection-symmetry-breaking solution ($Q_3 \neq 0$). Different colors correspond to $g_0'$ values as indicated in panel (b). Solid and dashed lines represent the $K=1$ and $K=0$ modes, respectively.}
\end{figure}
\begin{figure}[tbp]
\includegraphics[width=8.5cm]{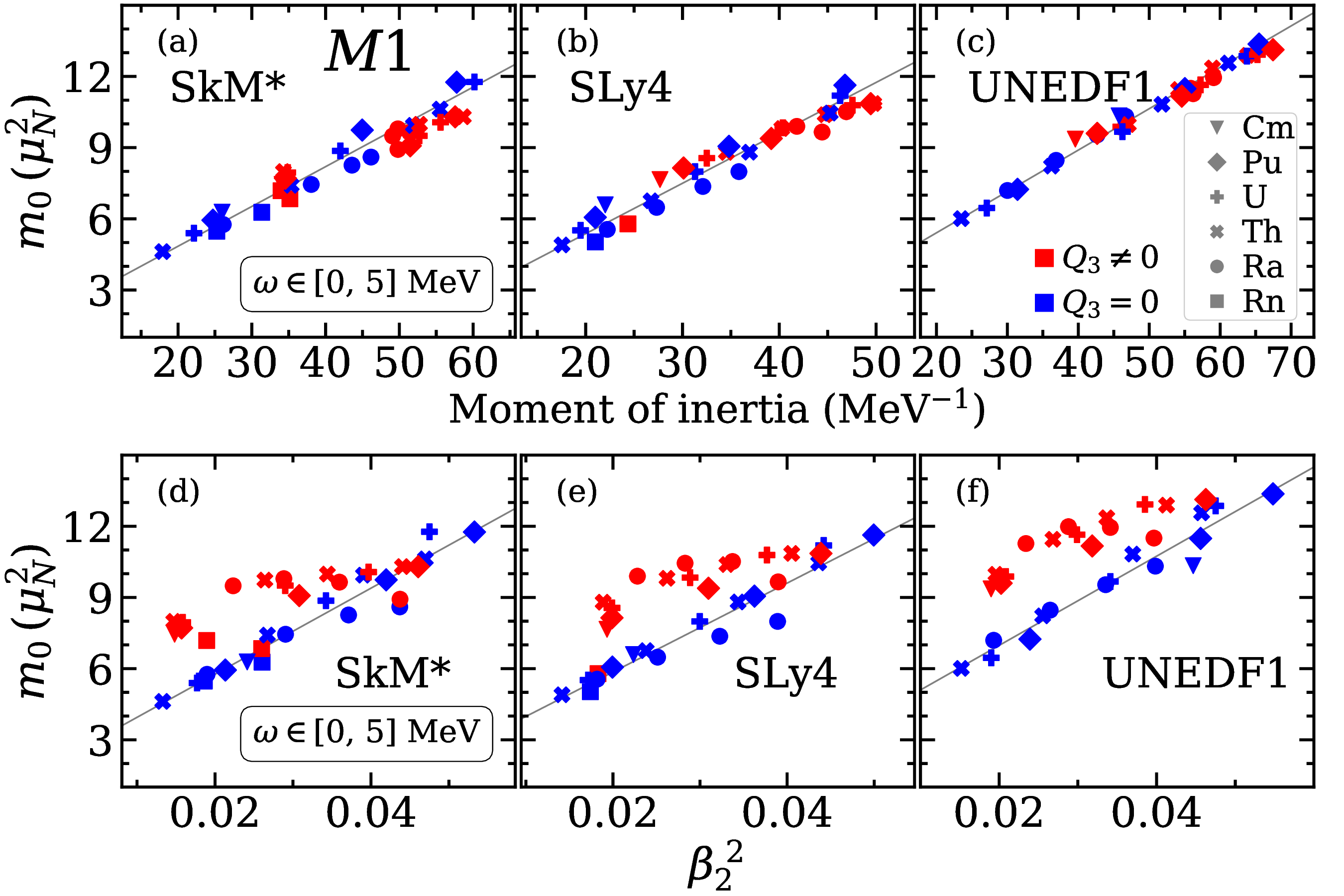}
\caption{\label{fig:M1_m0_vs_TV-MoI_beta2}Sum rules of magnetic dipole ($M1$) transitions for studied nuclei with $A=222\,\text{--}\,230$ using three different Skyrme functionals and including transitions with excitation energies between $0$ and \SI{5}{MeV}. Panels (a)--(c) show $m_0$ as a function of the Thouless--Valatin moment of inertia, while panels (d)--(f) show $m_0$ as a function of the square of the quadrupole deformation parameter $\beta_2$. Blue markers correspond to reflection-symmetry-conserving ($Q_3=0$) and red markers to reflection-symmetry-breaking ($Q_3\neq0$) HFB ground-state solutions. Different marker shapes indicate isotopic chains, and the straight line shows a linear fit to the $Q_3=0$ data.}
\end{figure}
\begin{figure*}[t]
\includegraphics[width=\textwidth]{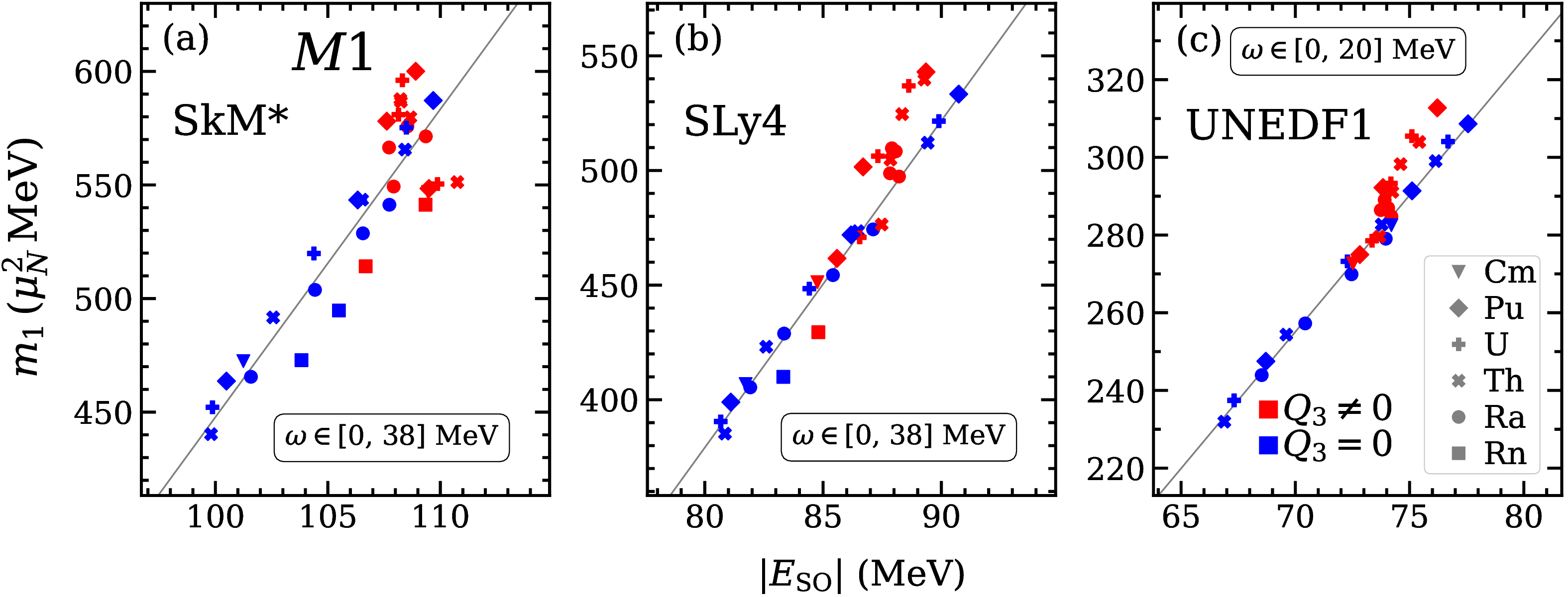}
\caption{\label{fig:M1_m1_vs_E_SO}Energy-weighted sum rules of magnetic dipole ($M1$) transitions as a function of the total spin-orbit energy $E_\text{SO}$ for studied nuclei with $A=222\,\text{--}\,230$. Panels show results for (a) SkM*, (b) SLy4, and (c) UNEDF1 functionals, including transitions with excitation energies as indicated in the panels for each Skyrme functional. Otherwise, results are displayed using the same conventions as in Fig.~\ref{fig:M1_m0_vs_TV-MoI_beta2}.}
\end{figure*}
Before proceeding to a more quantitative discussion of the $M1$ transition strengths, we recall that $M1$ properties can be sensitive to the time-odd part of the Skyrme functional (see, e.g., Ref.~\cite{Sassarini_2022}). The time-odd part, in turn, is generally not well constrained by the ground-state observables of the even-even nuclei used in the fitting protocols of the corresponding functionals, a limitation that also applies to the SkM* and SLy4 functionals employed in the present work. In contrast, the time-odd part of the UNEDF1 functional, as employed in the present work, is adjusted \cite{Sassarini_2022} to the magnetic dipole moments of next-to-doubly-magic odd-A nuclei. However, as noted in Ref.~\cite{Dobaczewski2026}, including data of deformed open-shell nuclei in its adjustment procedure, along with other improvements in the functional, could benefit its description of $M1$ moments, and in this regard, some uncertainty also remains for our $M1$ transition strengths of this functional. On the other hand, the qualitative observations above are consistent across the three Skyrme functionals employed (which all differ in their coupling constants of the time-odd Skyrme functional) with respect to the impact of octupole deformation on the $M1$ transition strength functions. However, to further assess the uncertainties from the time-odd sector, we have performed a series of $M1$ calculations for $^{224}\text{U}$ using UNEDF1 with a varying Landau parameter $g_0'$ within the uncertainty range reported in Ref.~\cite{Sassarini_2022}. Specifically, we carried out calculations for values of $g_0'=$ 1.3, 1.5, 1.7, 1.9, and 2.1. The corresponding results, shown in Fig.~\ref{fig:M1_Landau}, indicate that varying $g_0'$ has only a moderate effect on the transition strength functions. Transition strengths are mostly decreased in the spin-flip region with increasing $g_0'$, but most importantly, the fragmentation patterns of both the reflection-symmetric and reflection-symmetry-breaking solutions remain unchanged over the entire studied excitation energy range, regardless of the value of $g_0'$ used. Therefore, based on the consistency across the employed Skyrme functionals and the weak dependence on variations in the Landau parameter $g_0'$, we conclude that the observations on the impact of octupole deformation on the $M1$ transition strengths do not appear particularly sensitive to uncertainties in the time-odd part of the Skyrme functional.

To obtain a more quantitative measure of the differences in transition strengths between reflection-symmetric and reflection-symmetry-breaking HFB solutions, we calculated the corresponding $m_0$ sum rules, as given by Eq.~(\ref{eq:m_k}), for the low-energy region ($0\,\text{--}\,\SI{5}{MeV}$). Since this low-energy summed transition strength is expected to depend linearly on the moment of inertia, as shown by the two-rotor model (TRM) and by other approaches \cite{Lo_Iudice1997, Heyde2010}, we present the $m_0$ values as a function of the Thouless--Valatin moment of inertia (\ref{eq:TV_MoI}). Figures~\ref{fig:M1_m0_vs_TV-MoI_beta2}(a)--\ref{fig:M1_m0_vs_TV-MoI_beta2}(c) present the sum rules for all studied nuclei with mass numbers $A = 222\,\text{--}\,230$ and show both types of deformed HFB solutions following the same linear dependence on the moment of inertia. By contrast, the octupole-deformed solutions tend to exhibit a smaller spread in moment-of-inertia values, which are clustered toward the higher end of the obtained values. Therefore, the increased $M1$ transition strengths at low energies may be linked to the larger moment of inertia in the octupole-deformed solutions.

Figures~\ref{fig:M1_m0_vs_TV-MoI_beta2}(d)--\ref{fig:M1_m0_vs_TV-MoI_beta2}(f) show the $m_0$ values as a function of the squared quadrupole deformation parameter $\beta_2$ to examine the known deformation law \cite{Heyde2010, Lo_Iudice1997, Ziegler1990}, according to which the corresponding sum rule is quadratically correlated with the quadrupole deformation parameter. In contrast to the linear dependence on the moment of inertia, the deformation law appears to be violated by the octupole-deformed HFB solutions: while the reflection-symmetric solution shows a clear quadratic correlation with $\beta_2$, the octupole-deformed solutions, despite exhibiting a similar overall trend, display a larger spread and deviate from the quadratic dependence observed in the reflection-symmetric case. However, as shown in the Supplemental Material \cite{Supplemental_Material}, the increase in $m_0$ values for the octupole-deformed solutions does not correlate with the magnitude of the octupole deformation. These observations suggest that octupole deformation is not the direct cause of the enhanced sum rules.

When magnetic dipole transitions are considered over a wider energy range, the energy-weighted sum rule $m_1$ becomes a more relevant measure of the summed transition strength. Therefore, we calculated $m_1$ by integrating the energy-weighted transition strengths over the entire energy range considered for each Skyrme functional. Because the energy-weighted sum rule can, in some cases, be approximated \cite{Kurath1963} from the total spin-orbit energy of the ground-state solution with a linear dependence on $E_\text{SO}$ \cite{Kurath1963, Tong2024}, we plotted the $m_1$ values for all studied nuclei with $A = 222\,\text{--}\,230$ as a function of $E_\text{SO}$ in Fig.~\ref{fig:M1_m1_vs_E_SO}. The expected linear dependence of $m_1$ on $E_\text{SO}$ is observed to hold rather well for both types of deformed ground-state solutions. The same figure also shows that the octupole-deformed solutions often have a larger spin-orbit energy $|E_\text{SO}|$, similarly to their larger moments of inertia. Because the differences in the $M1$ transition strength functions between the two deformed ground-state solutions are concentrated at low energies, the larger total spin-orbit energies could also explain the enhanced transition strengths and $m_0$ values at low excitation energies in the octupole-deformed solutions.

\begin{figure}[t]
\includegraphics[width=8.5cm]{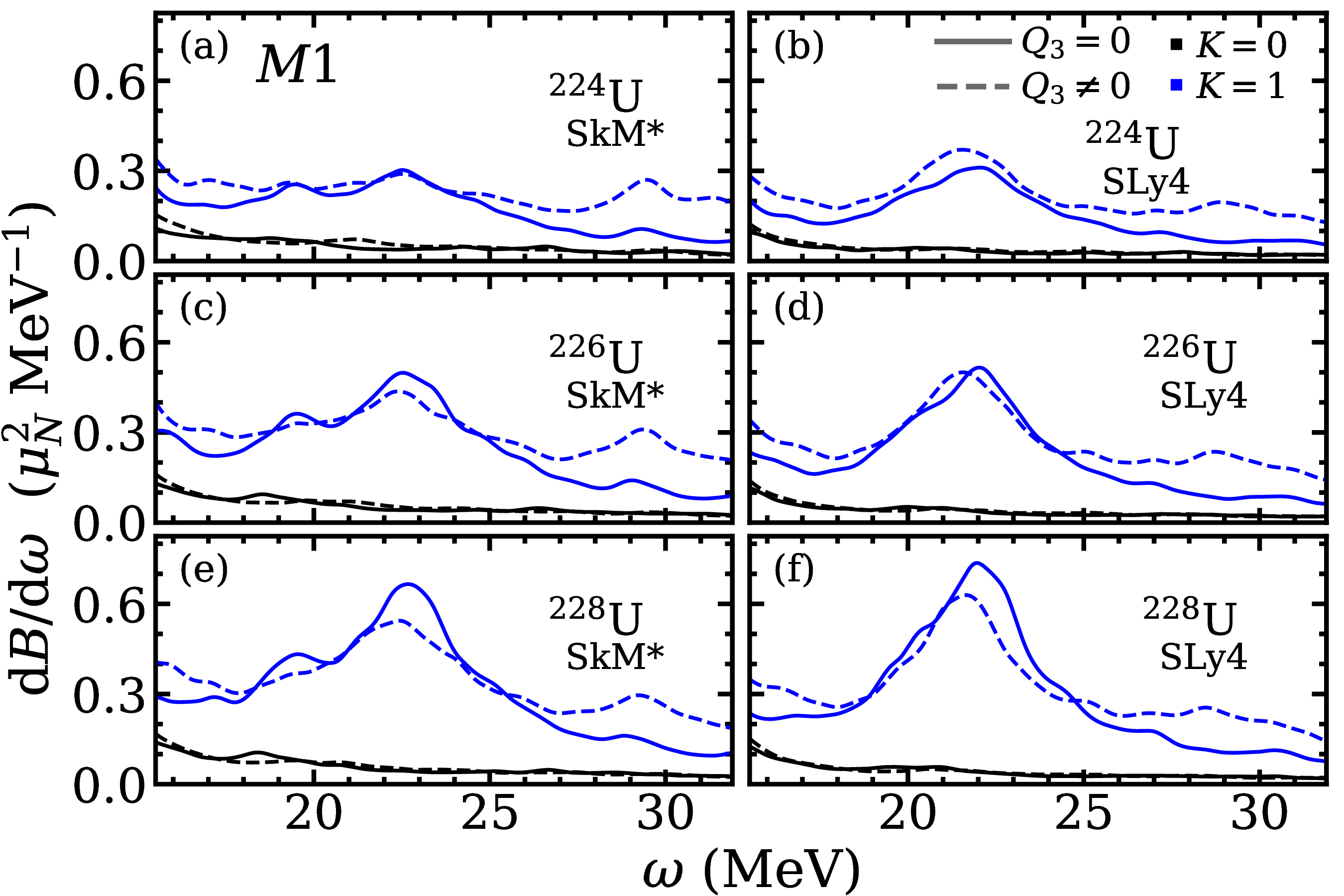}
\caption{\label{fig:M1_TS_high_energy_SR}Same as Fig.~\ref{fig:M1_U}, but showing only the SkM* and SLy4 functionals focusing on the higher excitation energies.}
\end{figure}
\begin{figure}[t]
\includegraphics[width=8.5cm]{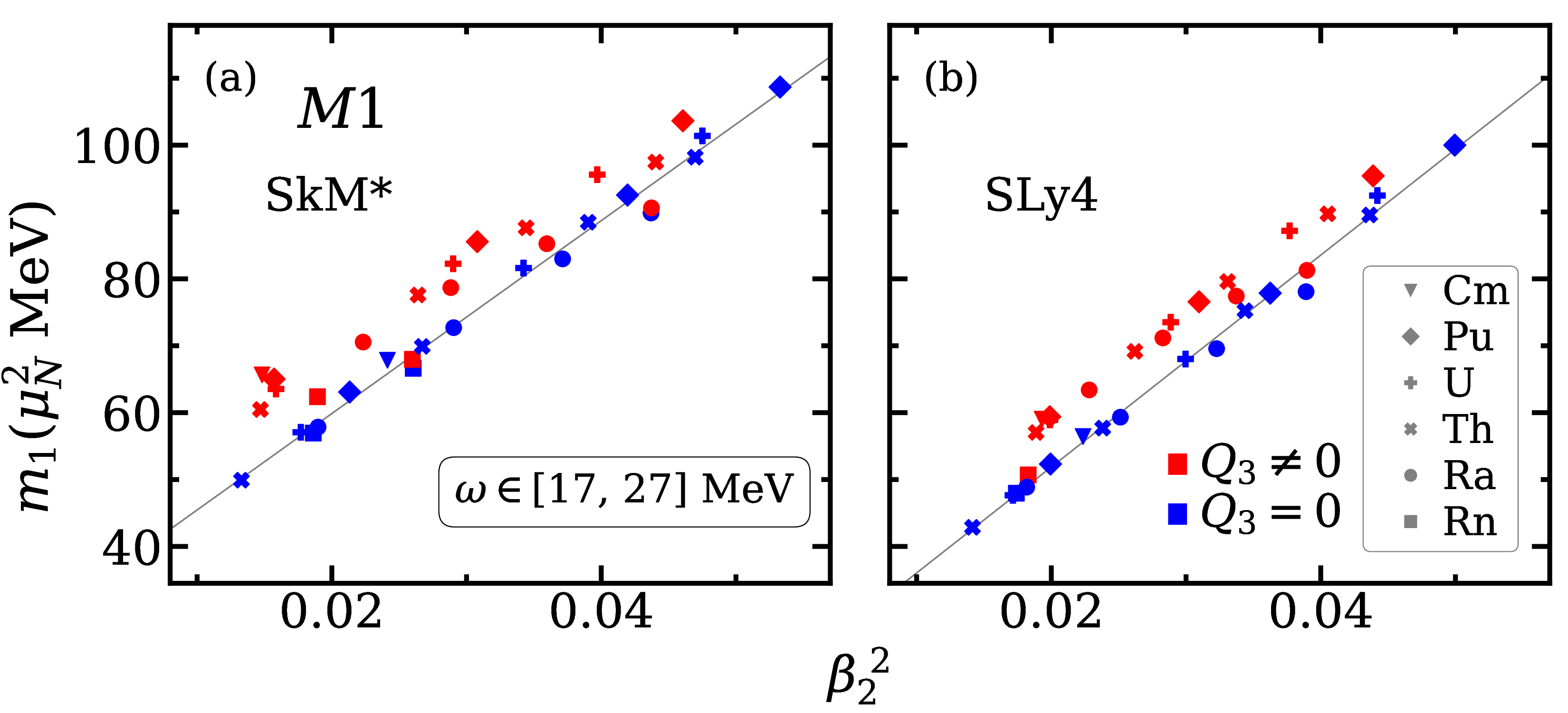}
\caption{\label{fig:M1_m1_vs_beta2_high_energy_SR}Energy-weighted sum rules $m_1$ of magnetic dipole ($M1$) transitions as a function of the square of the quadrupole deformation parameter $\beta_2$ for the studied nuclei with $A=222\,\text{--}\,230$. Panels show results for (a) SkM* and (b) SLy4 functionals, including transitions with excitation energies between $17$ and $\SI{27}{MeV}$. Otherwise, results are displayed using the same conventions as in Fig.~\ref{fig:M1_m0_vs_TV-MoI_beta2}.}
\end{figure}
As a final remark on $M1$ transitions, we consider the high-energy scissors mode, which could stem from the $K=1$ component of the isovector $E2$ mode \cite{Heyde2010}; due to nuclear deformation, $L$ is not a good quantum number, allowing quadrupole and magnetic dipole modes to mix. This mode has been found by schematic and realistic RPA models, with excitation energy of about $\SI{23}{MeV}$ for nuclei with $A=224$ \cite{Lo_Iudice2000}. Figure~\ref{fig:M1_TS_high_energy_SR} shows the transition strength functions of the U isotopes, focusing on the high-excitation-energy region for SkM* and SLy4 functionals. Although the transition strengths of the high-energy scissors mode are much smaller than those of the low-energy mode or spin-flip transitions, a clear peak appears near the expected excitation energy in the heavier, well-deformed isotopes. Since this scissors mode should also follow the deformation law, we present in Fig.~\ref{fig:M1_m1_vs_beta2_high_energy_SR} the $m_1$ sum rules as a function of the squared deformation parameter $\beta_2$ for all studied nuclei with $A = 222\,\text{--}\,230$, accounting for transitions from $17$ to $\SI{27}{MeV}$. Figure~\ref{fig:M1_m1_vs_beta2_high_energy_SR} shows that, in this case, the quadratic dependence of $m_1$ on $\beta_2$ holds regardless of the octupole deformation. Additionally, Fig.~\ref{fig:M1_m1_vs_beta2_high_energy_SR} shows that the octupole-deformed solutions have slightly larger $m_1$ values for this energy range compared to reflection-symmetric solutions, which could result from their larger background transition strengths, as suggested by Fig.~\ref{fig:M1_TS_high_energy_SR}. In turn, the choice of energy limits for the considered transitions is somewhat arbitrary, and for a smaller energy window, the increase in $m_1$ values for the octupole-deformed solutions disappears as shown in the Supplemental Material \cite{Supplemental_Material}.

\begin{figure*}[t]
    \includegraphics[width=\textwidth]{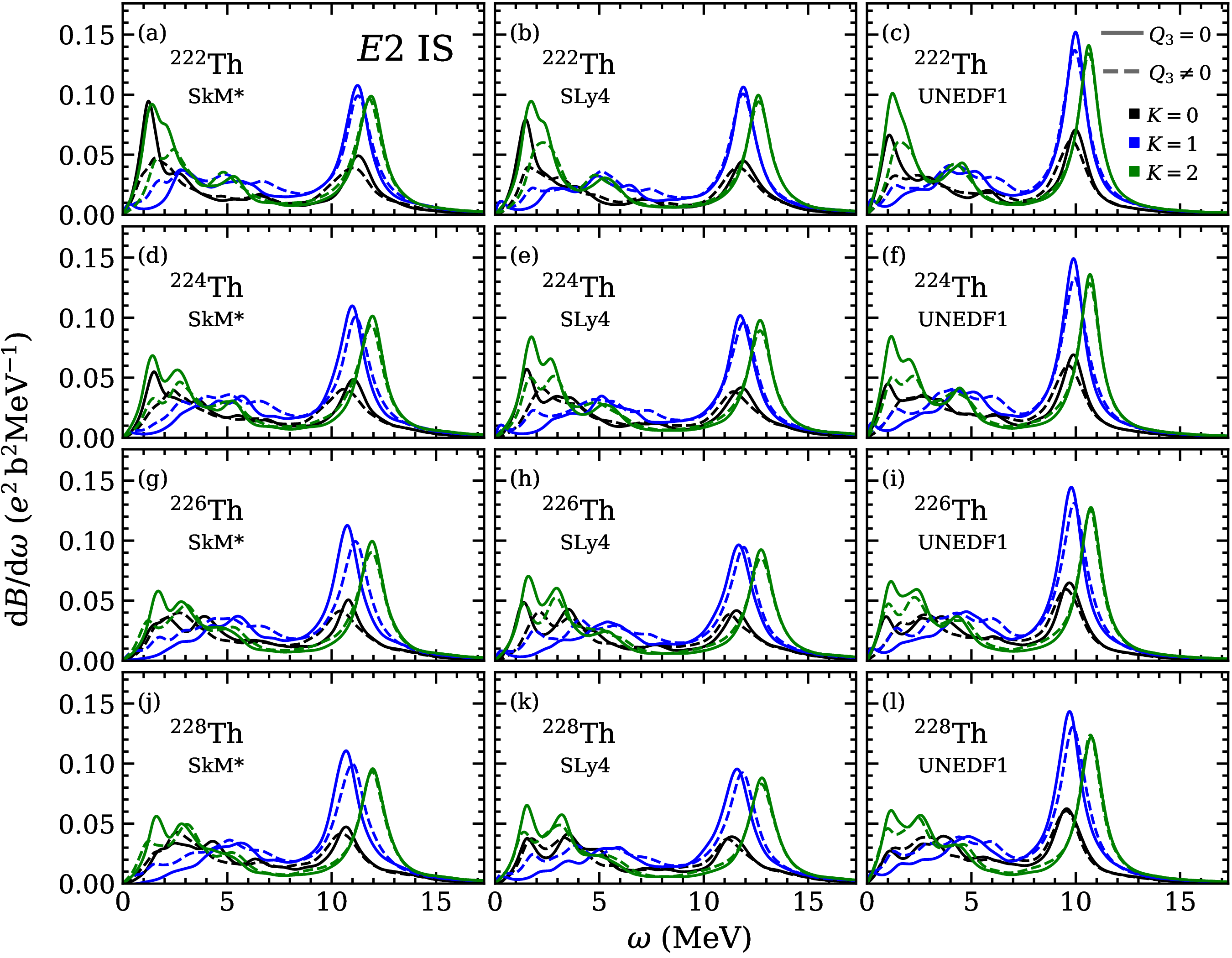}
    \caption{\label{fig:E2_IS}Same as Fig.~\ref{fig:M1_U}, but for isoscalar (IS) electric quadrupole ($E2$) transition strength functions of the studied Th isotopes.}
    \end{figure*}
\subsection{\textit{E}2 transitions}
\label{sec:E2-transitions}
Figure~\ref{fig:E2_IS} shows the isoscalar $E2$ transition strength functions for Th isotopes, calculated with the three Skyrme functionals used in this work. In $E2$ transitions, the position of the resonance peak is uniquely related to the isoscalar effective mass of the Skyrme parametrization. The best agreement with experimental data is known to be obtained for $m_0^*/m\!\approx\!0.78$, whereas smaller or larger values shift the resonance peak to too high or too low energies, respectively \cite{Reinhard1999}. Consistent with this trend, the resonance peaks appear at approximately $\SI{10}{MeV}$ for UNEDF1 ($m_0^*/m=0.99$ \cite{Kortelainen2012}), $\SI{11}{MeV}$ for SkM* ($m_0^*/m=0.79$ \cite{Bartel1982}), and $\SI{12}{MeV}$ for SLy4 ($m_0^*/m=0.70$ \cite{Chabanat1998}).

In the Th isotopes shown in Fig.~\ref{fig:E2_IS}, as well as in other nuclei presented in the Supplemental Material \cite{Supplemental_Material}, small differences in transition strengths at the resonances appear rather consistently between the two deformations across all $K$ modes. For the $K=0$ mode, the resonance peak is generally smaller and spread toward lower energies when the HFB solution is octupole deformed, particularly with the SkM* functional. The $K=1$ and $K=2$ peaks can also be slightly reduced for octupole-deformed solutions. Moreover, in the $K=1$ mode, the peak is shifted to slightly higher energy, while the $K=2$ peak remains at the same energy.

Isoscalar transition strength functions also show larger values at low energies. In general, the excitation spectrum is not expected to be continuous at low energies; therefore, the displayed functions do not represent the actual spectra very well. However, they can give some indications about the excitation modes. Larger peaks in the $K=0$ mode indicate softness of the PES with respect to axially symmetric quadrupole deformation $Q_{20}$. The octupole-deformed HFB solution often shows a much smaller $K=0$ peak, which could be due to the lowered energy of the solution, possibly making the PES steeper in the $Q_{20}$ direction. Larger low-energy peaks in the $K=2$ mode can be attributed to possible $\gamma$ softness of the PES. A large peak below $\SI{1}{MeV}$ could also indicate an unstable QRPA mode with imaginary excitation energy, implying that the found ground-state solution is a saddle point with respect to $\gamma$ deformation. This means that the ground state could be triaxial, at least if reflection symmetry breaking is not allowed. Such cases were recently further investigated in Ref.~\cite{Tong2024} by studying the PES in the ($\beta$, $\gamma$) plane. For the $K=1$ mode, the rotational spurious mode was removed; thus, there are no large low-energy transition strengths associated with it.

\begin{figure}[t]
\includegraphics[width=8.5cm]{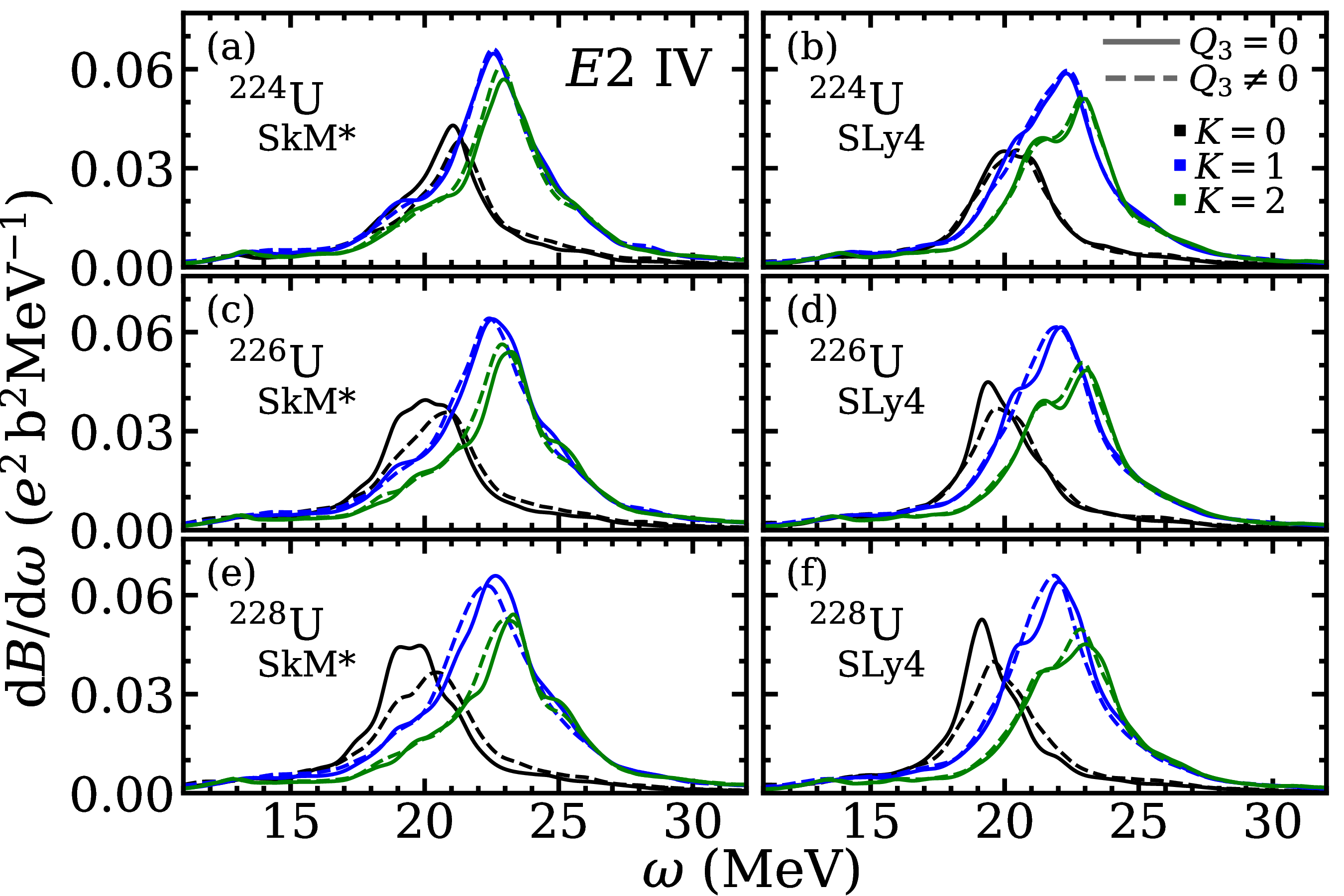}
\caption{\label{fig:E2_IV}Same as Fig.~\ref{fig:M1_U}, but for isovector (IV) electric quadrupole ($E2$) transition strength functions of the studied U isotopes, calculated with two Skyrme EDFs (SkM* and SLy4), as indicated in each panel.}
\end{figure}
Figure~\ref{fig:E2_IV} shows isovector $E2$ transition strength functions for U isotopes calculated with the SkM* and SLy4 functionals. The transition strength is mostly distributed between $17$ and $\SI{25}{MeV}$, so for that reason, calculations with the UNEDF1 functional were not performed, due to its convergence issues above $\SI{20}{MeV}$. For isovector $E2$ transitions, the strength functions of the $K=1$ and $K=2$ modes are very similar between reflection-symmetric and reflection symmetry-breaking HFB solutions. In contrast, the $K=0$ transition strength functions differ between the $Q_3=0$ and $Q_3 \neq 0$ cases, as the peak tends to be reduced and slightly shifted to higher energy in the octupole-deformed HFB solution. The differences are similar for both Skyrme functionals and across all isotopes studied.

\subsection{\textit{E}3 transitions}
\label{sec:E3-transitions}
\begin{figure}[t]
\includegraphics[width=8.5cm]{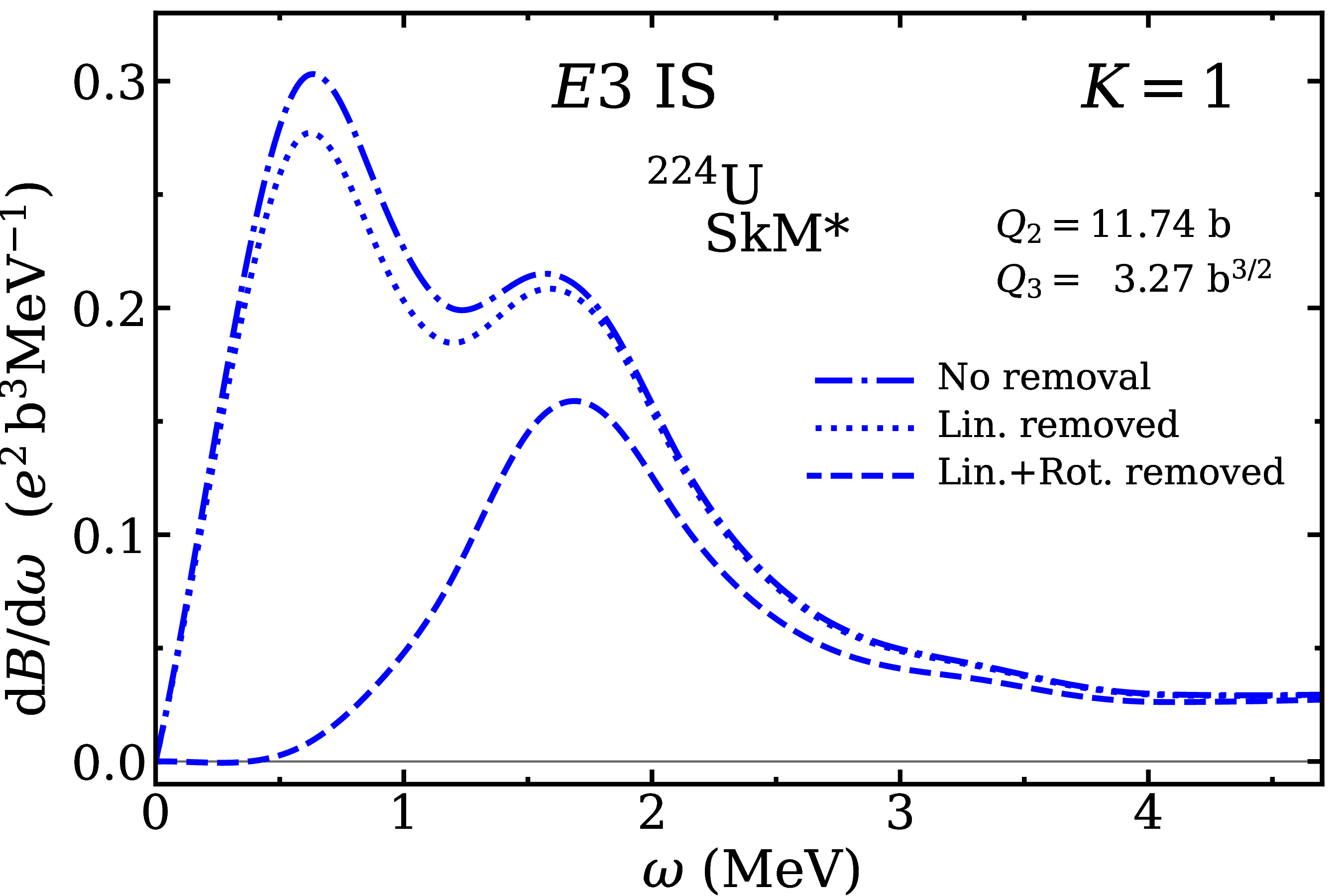}
\caption{\label{fig:E3_IS_spurious_removal}Isoscalar (IS) electric octupole ($E3$) transition strength functions of the $K=1$ mode (shown as the sum of the identical $K=\pm1$ modes) for the $^{224}\text{U}$ isotope. The strength functions are calculated on top of the octupole-deformed ground-state solution with given quadrupole ($Q_2$) and octupole ($Q_3$) moments, using the SkM* functional. The curves correspond to different treatments of spurious Nambu--Goldstone (NG) modes: dash-dotted line presents no removal, dotted line removal of the linear momentum spurious mode only (Lin. removed), and dashed line removal of both the linear and rotational momentum spurious modes (Lin. + Rot. removed).}
\end{figure}
$E3$ transition strengths can include spurious NG contributions in the $K=0$ and $K=1$ modes, which are related to the broken translational invariance. Additionally, when considering octupole-deformed HFB solutions with broken parity symmetry, the $K=1$ mode in the isoscalar transitions may also include a spurious NG mode associated with broken rotational symmetry. To investigate this possibility, we performed $E3$ IS FAM-QRPA calculations on top of the octupole-deformed ground-state solution using three different approaches for removing spurious modes: no removal, removal of only the linear momentum spurious mode, and removal of both the linear and rotational momentum spurious modes. Figure~\ref{fig:E3_IS_spurious_removal} shows the results of this comparison for the $^{224}\text{U}$ SkM* calculation, illustrating the effects of removing spurious modes at low excitation energies. When no spurious modes are removed, or only the linear spurious mode is removed, the transition strength function exhibits a clear two-peak structure, with the second peak having slightly lower transition strength values. When the rotational spurious mode is also removed, the two-peak structure disappears, and the transition strength function approaches zero below excitation energies of $\SI{1}{MeV}$. In Fig.~\ref{fig:E3_IS_spurious_removal}, the gray line at $\text{d}B/\text{d}\omega = 0$ indicates that in SkM* calculations, transition strengths become slightly negative at energies below $\SI{0.5}{MeV}$ when both spurious modes are removed. However, our analysis shows that in parity-breaking mean-field solutions, the rotational spurious mode can have a considerably more significant contribution to the calculated strength functions than the linear momentum spurious mode, and therefore, its removal should be considered essential.

\begin{figure}[t]
\includegraphics[width=8.5cm]{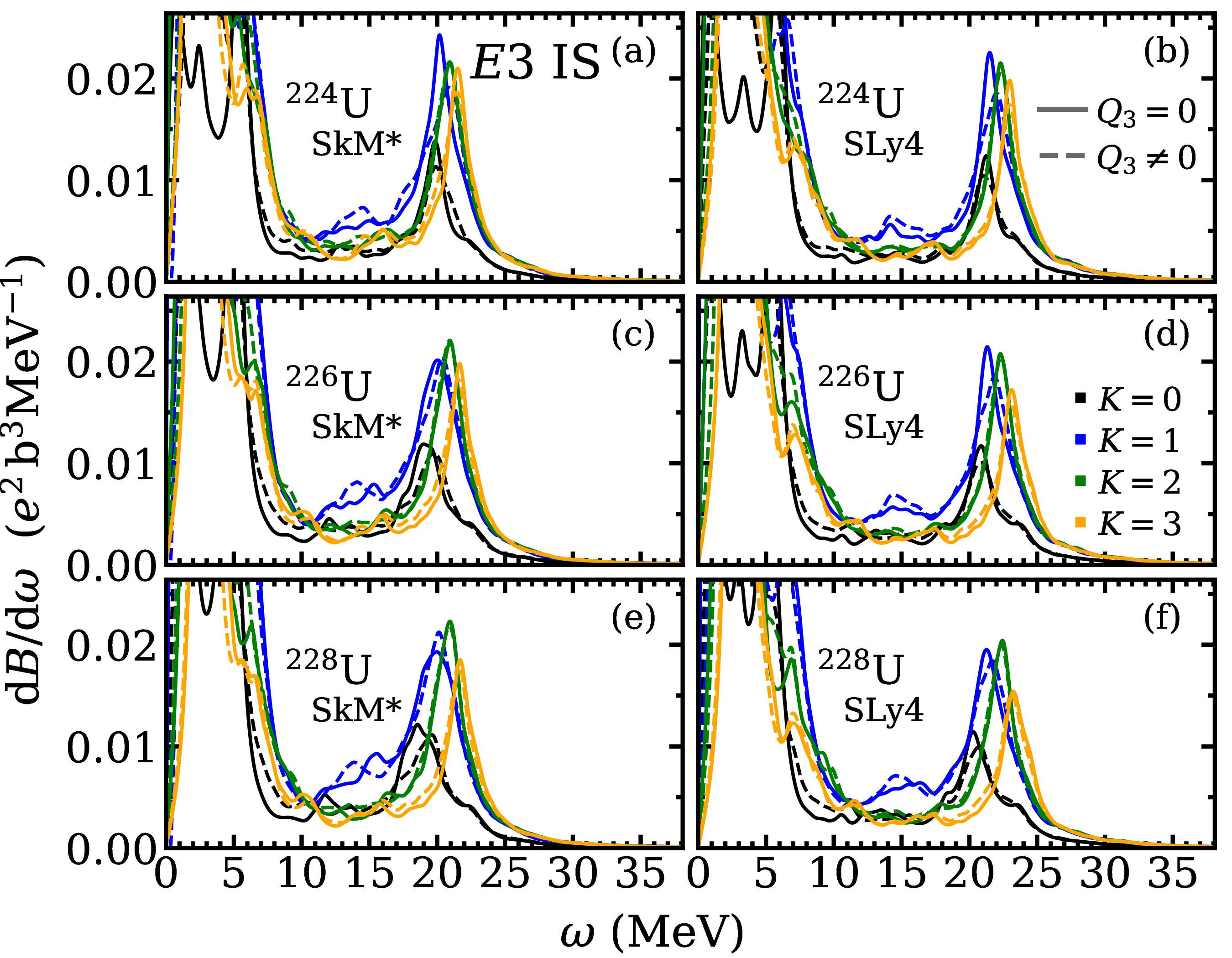}
\caption{\label{fig:E3_IS}Same as Fig.~\ref{fig:M1_U}, but for isoscalar (IS) electric octupole ($E3$) transition strength functions of the studied U isotopes, calculated with two Skyrme EDFs (SkM* and SLy4), as indicated in each panel.}
\end{figure}
\begin{figure}[t]
\includegraphics[width=8.5cm]{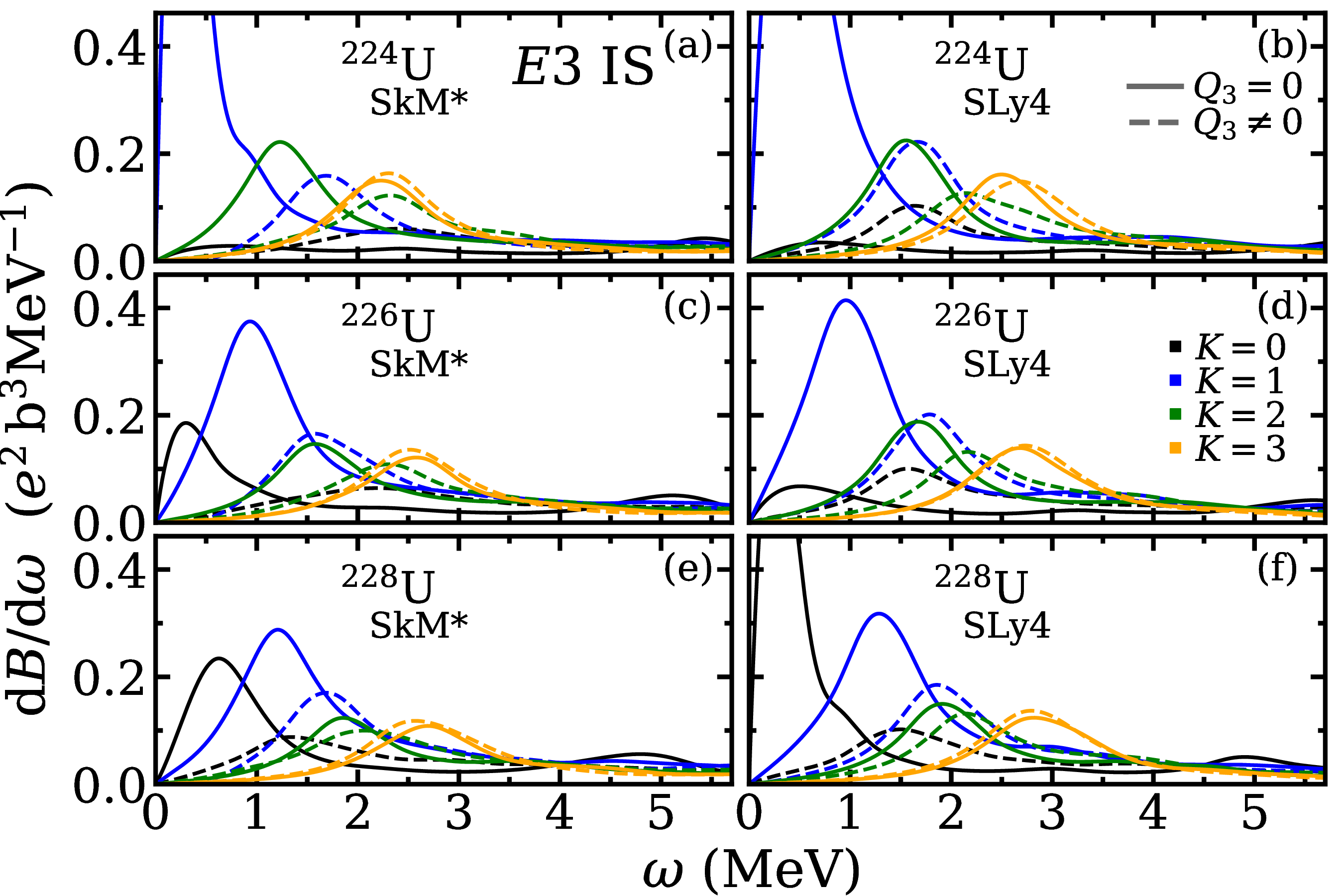}
\caption{\label{fig:E3_IS_low_energies}Same as Fig.~\ref{fig:E3_IS}, but focusing at the lower excitation energies.}
\end{figure}
Isoscalar $E3$ transition strength functions of U isotopes computed with SkM* and SLy4 functionals are presented in Figs.~\ref{fig:E3_IS} and \ref{fig:E3_IS_low_energies} for the full and the low-energy ranges, respectively. The removal of the NG mode, associated with the spurious center-of-mass motion, is performed for the $K=0$ and $K=1$ modes, while the removal of the rotational spurious mode is performed only for the $K=1$ mode calculated based on the octupole-deformed HFB solution. UNEDF1 calculations are omitted because significant transition strengths appear above $\SI{20}{MeV}$. The figures show that transition strengths at low energies are significantly larger than those at resonance energies. The most notable differences around resonances occur in the $K=0$ mode, where transition strengths for the octupole-deformed HFB solution are smaller than those for the reflection-symmetric solution. The largest differences for the $K=0$ mode are perhaps expected, as it corresponds to the operator relevant to octupole deformation. The $K=1$ mode shows some difference between the two HFB solutions, with the reflection-symmetric solution exhibiting a bit higher peak. Conversely, the $K=2$ and $K=3$ modes are very similar for both deformations.

At low energies in the isoscalar $E3$ transition strengths shown in Fig.~\ref{fig:E3_IS_low_energies}, two possible causes for the large transition strengths in each mode can be identified: softness of the PES or the presence of a saddle point, the latter resulting in an unstable QRPA mode associated with a pole at imaginary energy. Such an unstable mode is expected for the $K=0$ excitation of nonoctupole-deformed HFB solutions, as these solutions correspond to a saddle point of the PES along the $Q_{30}$ direction. Indeed, the nonoctupole-deformed solution sometimes exhibits larger transition strengths very close to zero energy. The $K=1$ mode also exhibits larger transition strengths at low energies for both deformations, particularly for the parity-conserving HFB solution. This suggests that at the reflection-symmetric solution, the PES is very soft or even a saddle point with respect to the $Q_{31}$ direction. Additionally, the $K=2$ and $K=3$ modes show larger low-energy transition strengths; however, these are smaller and occur at higher energies compared to the $K=0$ and $K=1$ modes, indicating a less soft PES in the $Q_{32}$ and $Q_{33}$ directions.

\begin{figure}[t]
\includegraphics[width=8.5cm]{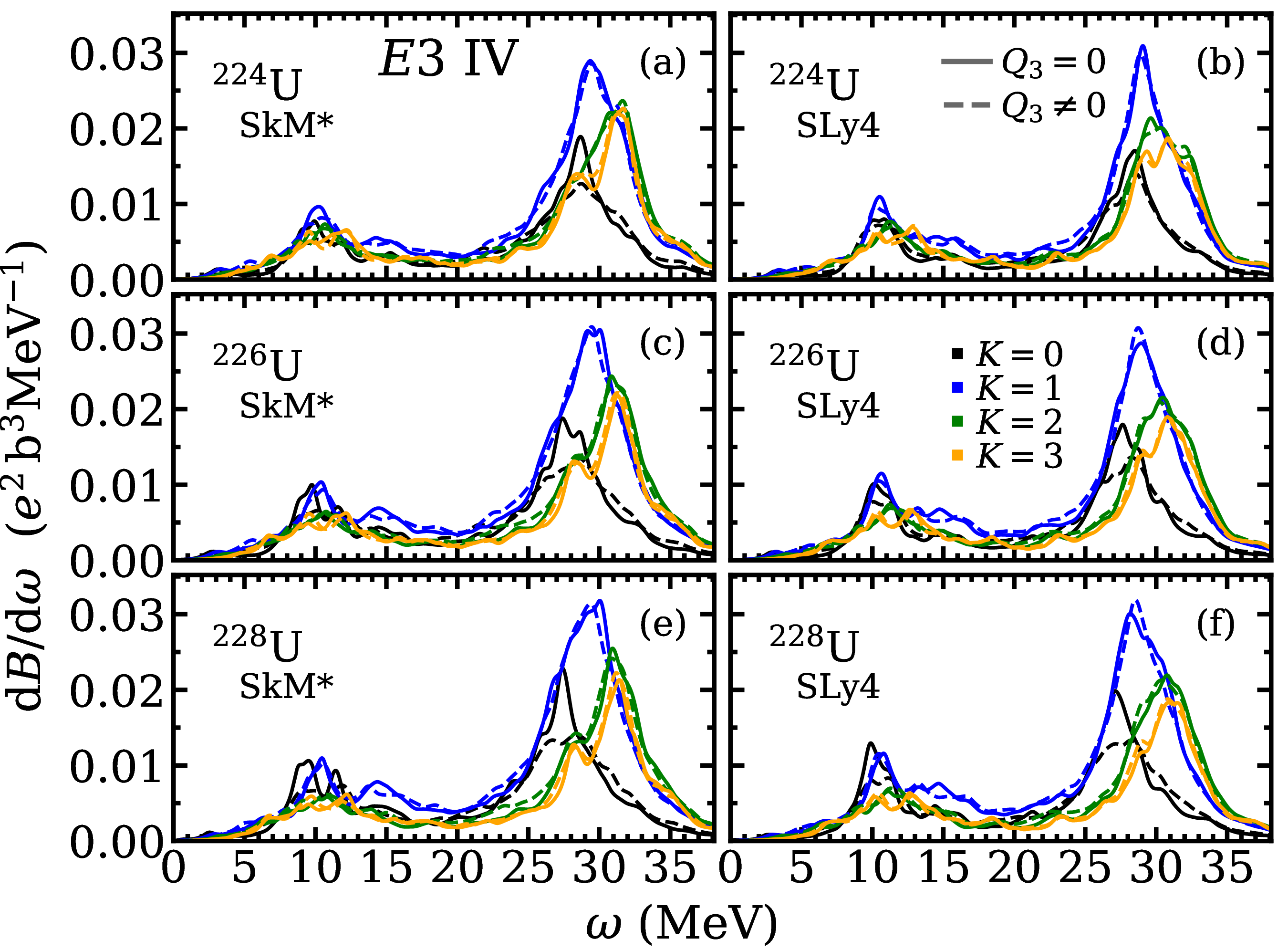}
\caption{\label{fig:E3_IV}Same as Fig.~\ref{fig:M1_U}, but for isovector (IV) electric octupole ($E3$) transition strength functions of the studied U isotopes, calculated with two Skyrme EDFs (SkM* and SLy4), as indicated in each panel.}
\end{figure}
Figure~\ref{fig:E3_IV} shows the isovector $E3$ transition strength functions of U isotopes computed with the SkM* and SLy4 functionals over the full calculated energy range. Again, UNEDF1 calculations are omitted due to the significant portion of transition strengths at energies above $\SI{20}{MeV}$. In these calculations, the spurious center-of-mass motion was removed from the $K=0$ and $K=1$ modes. Some differences between the two deformations can be observed, particularly for the $K=0$ mode, resembling those described in the isoscalar results. In contrast, the $K=1$, $K=2$, and $K=3$ modes show practically no difference between the two deformed HFB ground-state solutions.

\section{Conclusions}
\label{sec:Conclusions}
In this work, we calculated the electric and magnetic multipole responses of actinide nuclei with a predicted octupole-deformed ground state to investigate the impact of octupole deformation on the corresponding transition strengths. The analysis was carried out using the iterative FAM-QRPA method based on two distinct deformed Skyrme--HFB ground-state solutions: one constrained to conserve reflection symmetry, and another allowing the breaking of reflection symmetry. Calculations were performed using three different Skyrme functionals: SkM*, SLy4, and UNEDF1.

We found that octupole deformation generally has only a modest impact on the transition strengths at the resonances for the electric dipole, quadrupole, and octupole modes. At low excitation energies, however, the isoscalar electric quadrupole and octupole transition strength functions were noticeably affected by octupole deformation compared to the reflection-symmetric HFB solution. This behavior can be understood in terms of changes in the shape of the PES between the positions corresponding to these two deformed HFB solutions.

Meanwhile, the magnetic dipole transition strength functions showed clear differences between the reflection-symmetry-conserving and -breaking HFB solutions in the energy range $0\,\text{--}\,\SI{8}{MeV}$. The observed increase in the summed transition strength in this energy range could be related to the higher moment of inertia and total spin-orbit energy of the octupole-deformed HFB solutions, rather than to octupole deformation itself. However, octupole-deformed HFB solutions were found to violate the low-energy deformation law of the magnetic dipole transition sum rules, which could serve as an additional signature of octupole deformation in experimental studies.

In our analysis, the $K=1^-$ mode was found to contain a significant contribution from the rotational spurious mode in the isoscalar $E3$ transition strength function calculated using the octupole-deformed HFB solutions. This can be attributed to the non-conservation of parity in octupole-deformed solutions. Since this contribution was so significant and much larger than that from the linear-momentum spurious mode, it was considered essential to remove it in such cases.

As a continuation of this work, a detailed comparison of individual low-energy excitation modes in the $M1$ spectra of octupole-deformed and non-octupole-deformed solutions would be particularly intriguing. Such an investigation could be carried out within the matrix-formulated QRPA framework or by applying the techniques introduced in Ref.~\cite{Hinohara2013} to the FAM-QRPA approach. Furthermore, assessing the impact of parity restoration -- with or without the angular momentum projection (see recent study in Ref.~\cite{Porro2024}) -- on the transition strengths of the octupole-deformed solution would be a natural extension of this work.

\begin{acknowledgments}

This work has been supported by the Research Council of Finland (Project No. 339243 and the Centre of Excellence in Neutron-Star Physics, Project No. 374066) and by funding from the doctoral training pilot. We acknowledge the CSC-IT Center for Science Ltd., Finland, for the allocation of computational resources.
\end{acknowledgments}

\section*{Data availability}
The data that support the findings of this article are openly available \cite{Data_repository}.

\providecommand{\noopsort}[1]{}\providecommand{\singleletter}[1]{#1}%


\begin{thebibliography}{64}%
\makeatletter
\providecommand \@ifxundefined [1]{%
 \@ifx{#1\undefined}
}%
\providecommand \@ifnum [1]{%
 \ifnum #1\expandafter \@firstoftwo
 \else \expandafter \@secondoftwo
 \fi
}%
\providecommand \@ifx [1]{%
 \ifx #1\expandafter \@firstoftwo
 \else \expandafter \@secondoftwo
 \fi
}%
\providecommand \natexlab [1]{#1}%
\providecommand \enquote  [1]{``#1''}%
\providecommand \bibnamefont  [1]{#1}%
\providecommand \bibfnamefont [1]{#1}%
\providecommand \citenamefont [1]{#1}%
\providecommand \href@noop [0]{\@secondoftwo}%
\providecommand \href [0]{\begingroup \@sanitize@url \@href}%
\providecommand \@href[1]{\@@startlink{#1}\@@href}%
\providecommand \@@href[1]{\endgroup#1\@@endlink}%
\providecommand \@sanitize@url [0]{\catcode `\\12\catcode `\$12\catcode `\&12\catcode `\#12\catcode `\^12\catcode `\_12\catcode `\%12\relax}%
\providecommand \@@startlink[1]{}%
\providecommand \@@endlink[0]{}%
\providecommand \url  [0]{\begingroup\@sanitize@url \@url }%
\providecommand \@url [1]{\endgroup\@href {#1}{\urlprefix }}%
\providecommand \urlprefix  [0]{URL }%
\providecommand \Eprint [0]{\href }%
\providecommand \doibase [0]{https://doi.org/}%
\providecommand \selectlanguage [0]{\@gobble}%
\providecommand \bibinfo  [0]{\@secondoftwo}%
\providecommand \bibfield  [0]{\@secondoftwo}%
\providecommand \translation [1]{[#1]}%
\providecommand \BibitemOpen [0]{}%
\providecommand \bibitemStop [0]{}%
\providecommand \bibitemNoStop [0]{.\EOS\space}%
\providecommand \EOS [0]{\spacefactor3000\relax}%
\providecommand \BibitemShut  [1]{\csname bibitem#1\endcsname}%
\let\auto@bib@innerbib\@empty
%</preamble>
\bibitem [{\citenamefont {Ring}\ and\ \citenamefont {Schuck}(1980)}]{Ring1980}%
  \BibitemOpen
  \bibfield  {author} {\bibinfo {author} {\bibfnamefont {P.}~\bibnamefont {Ring}}\ and\ \bibinfo {author} {\bibfnamefont {P.}~\bibnamefont {Schuck}},\ }\href@noop {} {\emph {\bibinfo {title} {The Nuclear Many-Body Problem}}}\ (\bibinfo  {publisher} {Springer-Verlag},\ \bibinfo {address} {New York},\ \bibinfo {year} {1980})\BibitemShut {NoStop}%
\bibitem [{\citenamefont {Bender}\ \emph {et~al.}(2003)\citenamefont {Bender}, \citenamefont {Heenen},\ and\ \citenamefont {Reinhard}}]{Bender2003}%
  \BibitemOpen
  \bibfield  {author} {\bibinfo {author} {\bibfnamefont {M.}~\bibnamefont {Bender}}, \bibinfo {author} {\bibfnamefont {P.-H.}\ \bibnamefont {Heenen}},\ and\ \bibinfo {author} {\bibfnamefont {P.-G.}\ \bibnamefont {Reinhard}},\ }\bibfield  {title} {\bibinfo {title} {Self-consistent mean-field models for nuclear structure},\ }\href {https://doi.org/10.1103/RevModPhys.75.121} {\bibfield  {journal} {\bibinfo  {journal} {Rev. Mod. Phys.}\ }\textbf {\bibinfo {volume} {75}},\ \bibinfo {pages} {121} (\bibinfo {year} {2003})}\BibitemShut {NoStop}%
\bibitem [{\citenamefont {Schunck}(2019)}]{EDF_book2019}%
  \BibitemOpen
  \bibinfo {editor} {\bibfnamefont {N.}~\bibnamefont {Schunck}},\ ed.,\ \href@noop {} {\emph {\bibinfo {title} {Energy Density Functional Methods for Atomic Nuclei}}}\ (\bibinfo  {publisher} {IOP Publishing},\ \bibinfo {address} {Bristol, UK},\ \bibinfo {year} {2019})\BibitemShut {NoStop}%
\bibitem [{\citenamefont {Hergert}(2020)}]{Hergert2020}%
  \BibitemOpen
  \bibfield  {author} {\bibinfo {author} {\bibfnamefont {H.}~\bibnamefont {Hergert}},\ }\bibfield  {title} {\bibinfo {title} {A guided tour of ab initio nuclear many-body theory},\ }\href {https://doi.org/10.3389/fphy.2020.00379} {\bibfield  {journal} {\bibinfo  {journal} {Front. Phys.}\ }\textbf {\bibinfo {volume} {8}},\ \bibinfo {pages} {379} (\bibinfo {year} {2020})}\BibitemShut {NoStop}%
\bibitem [{\citenamefont {Sheikh}\ \emph {et~al.}(2021)\citenamefont {Sheikh}, \citenamefont {Dobaczewski}, \citenamefont {Ring}, \citenamefont {Robledo},\ and\ \citenamefont {Yannouleas}}]{Sheikh2021}%
  \BibitemOpen
  \bibfield  {author} {\bibinfo {author} {\bibfnamefont {J.~A.}\ \bibnamefont {Sheikh}}, \bibinfo {author} {\bibfnamefont {J.}~\bibnamefont {Dobaczewski}}, \bibinfo {author} {\bibfnamefont {P.}~\bibnamefont {Ring}}, \bibinfo {author} {\bibfnamefont {L.~M.}\ \bibnamefont {Robledo}},\ and\ \bibinfo {author} {\bibfnamefont {C.}~\bibnamefont {Yannouleas}},\ }\bibfield  {title} {\bibinfo {title} {Symmetry restoration in mean-field approaches},\ }\href {https://doi.org/10.1088/1361-6471/ac288a} {\bibfield  {journal} {\bibinfo  {journal} {J. Phys. G:Nucl. Part. Phys.}\ }\textbf {\bibinfo {volume} {48}},\ \bibinfo {pages} {123001} (\bibinfo {year} {2021})}\BibitemShut {NoStop}%
\bibitem [{\citenamefont {Butler}\ and\ \citenamefont {Nazarewicz}(1996)}]{Butler1996}%
  \BibitemOpen
  \bibfield  {author} {\bibinfo {author} {\bibfnamefont {P.~A.}\ \bibnamefont {Butler}}\ and\ \bibinfo {author} {\bibfnamefont {W.}~\bibnamefont {Nazarewicz}},\ }\bibfield  {title} {\bibinfo {title} {Intrinsic reflection asymmetry in atomic nuclei},\ }\href {https://doi.org/10.1103/RevModPhys.68.349} {\bibfield  {journal} {\bibinfo  {journal} {Rev. Mod. Phys.}\ }\textbf {\bibinfo {volume} {68}},\ \bibinfo {pages} {349} (\bibinfo {year} {1996})}\BibitemShut {NoStop}%
\bibitem [{\citenamefont {Chen}\ \emph {et~al.}(2021)\citenamefont {Chen}, \citenamefont {Li}, \citenamefont {Dobaczewski},\ and\ \citenamefont {Nazarewicz}}]{Chen2021}%
  \BibitemOpen
  \bibfield  {author} {\bibinfo {author} {\bibfnamefont {M.}~\bibnamefont {Chen}}, \bibinfo {author} {\bibfnamefont {T.}~\bibnamefont {Li}}, \bibinfo {author} {\bibfnamefont {J.}~\bibnamefont {Dobaczewski}},\ and\ \bibinfo {author} {\bibfnamefont {W.}~\bibnamefont {Nazarewicz}},\ }\bibfield  {title} {\bibinfo {title} {Microscopic origin of reflection-asymmetric nuclear shapes},\ }\href {https://doi.org/10.1103/PhysRevC.103.034303} {\bibfield  {journal} {\bibinfo  {journal} {Phys. Rev. C}\ }\textbf {\bibinfo {volume} {103}},\ \bibinfo {pages} {034303} (\bibinfo {year} {2021})}\BibitemShut {NoStop}%
\bibitem [{\citenamefont {Butler}(2020)}]{Butler2020}%
  \BibitemOpen
  \bibfield  {author} {\bibinfo {author} {\bibfnamefont {P.~A.}\ \bibnamefont {Butler}},\ }\bibfield  {title} {\bibinfo {title} {Pear-shaped atomic nuclei},\ }\href {https://doi.org/10.1098/rspa.2020.0202} {\bibfield  {journal} {\bibinfo  {journal} {Proc. R. Soc. A}\ }\textbf {\bibinfo {volume} {476}},\ \bibinfo {pages} {20200202} (\bibinfo {year} {2020})}\BibitemShut {NoStop}%
\bibitem [{\citenamefont {Cao}\ \emph {et~al.}(2020)\citenamefont {Cao}, \citenamefont {Agbemava}, \citenamefont {Afanasjev}, \citenamefont {Nazarewicz},\ and\ \citenamefont {Olsen}}]{Cao2020}%
  \BibitemOpen
  \bibfield  {author} {\bibinfo {author} {\bibfnamefont {Y.}~\bibnamefont {Cao}}, \bibinfo {author} {\bibfnamefont {S.~E.}\ \bibnamefont {Agbemava}}, \bibinfo {author} {\bibfnamefont {A.~V.}\ \bibnamefont {Afanasjev}}, \bibinfo {author} {\bibfnamefont {W.}~\bibnamefont {Nazarewicz}},\ and\ \bibinfo {author} {\bibfnamefont {E.}~\bibnamefont {Olsen}},\ }\bibfield  {title} {\bibinfo {title} {Landscape of pear-shaped even-even nuclei},\ }\href {https://doi.org/10.1103/PhysRevC.102.024311} {\bibfield  {journal} {\bibinfo  {journal} {Phys. Rev. C}\ }\textbf {\bibinfo {volume} {102}},\ \bibinfo {pages} {024311} (\bibinfo {year} {2020})}\BibitemShut {NoStop}%
\bibitem [{\citenamefont {Agbemava}\ \emph {et~al.}(2016)\citenamefont {Agbemava}, \citenamefont {Afanasjev},\ and\ \citenamefont {Ring}}]{Agbemava2016}%
  \BibitemOpen
  \bibfield  {author} {\bibinfo {author} {\bibfnamefont {S.~E.}\ \bibnamefont {Agbemava}}, \bibinfo {author} {\bibfnamefont {A.~V.}\ \bibnamefont {Afanasjev}},\ and\ \bibinfo {author} {\bibfnamefont {P.}~\bibnamefont {Ring}},\ }\bibfield  {title} {\bibinfo {title} {{Octupole deformation in the ground states of even-even nuclei: A global analysis within the covariant density functional theory}},\ }\href {https://doi.org/10.1103/PhysRevC.93.044304} {\bibfield  {journal} {\bibinfo  {journal} {Phys. Rev. C}\ }\textbf {\bibinfo {volume} {93}},\ \bibinfo {pages} {044304} (\bibinfo {year} {2016})}\BibitemShut {NoStop}%
\bibitem [{\citenamefont {Ebata}\ and\ \citenamefont {Nakatsukasa}(2017)}]{Ebata2017}%
  \BibitemOpen
  \bibfield  {author} {\bibinfo {author} {\bibfnamefont {S.}~\bibnamefont {Ebata}}\ and\ \bibinfo {author} {\bibfnamefont {T.}~\bibnamefont {Nakatsukasa}},\ }\bibfield  {title} {\bibinfo {title} {{Octupole deformation in the nuclear chart based on the 3D Skyrme Hartree–Fock plus BCS model}},\ }\href {https://doi.org/10.1088/1402-4896/aa6c4c} {\bibfield  {journal} {\bibinfo  {journal} {Phys. Scr.}\ }\textbf {\bibinfo {volume} {92}},\ \bibinfo {pages} {064005} (\bibinfo {year} {2017})}\BibitemShut {NoStop}%
\bibitem [{\citenamefont {Grams}\ \emph {et~al.}(2023)\citenamefont {Grams}, \citenamefont {Ryssens}, \citenamefont {Scamps}, \citenamefont {Goriely},\ and\ \citenamefont {Chamel}}]{Grams2023}%
  \BibitemOpen
  \bibfield  {author} {\bibinfo {author} {\bibfnamefont {G.}~\bibnamefont {Grams}}, \bibinfo {author} {\bibfnamefont {W.}~\bibnamefont {Ryssens}}, \bibinfo {author} {\bibfnamefont {G.}~\bibnamefont {Scamps}}, \bibinfo {author} {\bibfnamefont {S.}~\bibnamefont {Goriely}},\ and\ \bibinfo {author} {\bibfnamefont {N.}~\bibnamefont {Chamel}},\ }\bibfield  {title} {\bibinfo {title} {{Skyrme-Hartree-Fock-Bogoliubov mass models on a 3D mesh: III. From atomic nuclei to neutron stars}},\ }\href {https://doi.org/10.1140/epja/s10050-023-01158-6} {\bibfield  {journal} {\bibinfo  {journal} {Eur. Phys. J. A}\ }\textbf {\bibinfo {volume} {59}},\ \bibinfo {pages} {270} (\bibinfo {year} {2023})}\BibitemShut {NoStop}%
\bibitem [{\citenamefont {Bohr}\ and\ \citenamefont {Mottelson}(1975)}]{Bohr1975}%
  \BibitemOpen
  \bibfield  {author} {\bibinfo {author} {\bibfnamefont {A.}~\bibnamefont {Bohr}}\ and\ \bibinfo {author} {\bibfnamefont {B.~R.}\ \bibnamefont {Mottelson}},\ }\href@noop {} {\emph {\bibinfo {title} {Nuclear Structure. Vol 2, Nuclear Deformations}}}\ (\bibinfo  {publisher} {Advanced Book Program, W.A.},\ \bibinfo {address} {Reading, Massachusetts},\ \bibinfo {year} {1975})\BibitemShut {NoStop}%
\bibitem [{\citenamefont {Roca-Maza}\ and\ \citenamefont {Paar}(2018)}]{Roca-Maza2018}%
  \BibitemOpen
  \bibfield  {author} {\bibinfo {author} {\bibfnamefont {X.}~\bibnamefont {Roca-Maza}}\ and\ \bibinfo {author} {\bibfnamefont {N.}~\bibnamefont {Paar}},\ }\bibfield  {title} {\bibinfo {title} {Nuclear equation of state from ground and collective excited state properties of nuclei},\ }\href {https://doi.org/https://doi.org/10.1016/j.ppnp.2018.04.001} {\bibfield  {journal} {\bibinfo  {journal} {Prog. Part. Nucl. Phys.}\ }\textbf {\bibinfo {volume} {101}},\ \bibinfo {pages} {96} (\bibinfo {year} {2018})}\BibitemShut {NoStop}%
\bibitem [{\citenamefont {Berman}\ and\ \citenamefont {Fultz}(1975)}]{Berman1975}%
  \BibitemOpen
  \bibfield  {author} {\bibinfo {author} {\bibfnamefont {B.~L.}\ \bibnamefont {Berman}}\ and\ \bibinfo {author} {\bibfnamefont {S.~C.}\ \bibnamefont {Fultz}},\ }\bibfield  {title} {\bibinfo {title} {Measurements of the giant dipole resonance with monoenergetic photons},\ }\href {https://doi.org/10.1103/RevModPhys.47.713} {\bibfield  {journal} {\bibinfo  {journal} {Rev. Mod. Phys.}\ }\textbf {\bibinfo {volume} {47}},\ \bibinfo {pages} {713} (\bibinfo {year} {1975})}\BibitemShut {NoStop}%
\bibitem [{\citenamefont {Hilton}()}]{Hilton1976}%
  \BibitemOpen
  \bibfield  {author} {\bibinfo {author} {\bibfnamefont {R.~R.}\ \bibnamefont {Hilton}},\ }\bibfield  {title} {\bibinfo {title} {Talk presented at the \emph{{International Conference on Nuclear Structure}}},\ }\bibinfo {note} {{JINR}, Dubna, (1976) (unpublished)}\BibitemShut {NoStop}%
\bibitem [{\citenamefont {Suzuki}\ and\ \citenamefont {Rowe}(1977)}]{Suzuki1977}%
  \BibitemOpen
  \bibfield  {author} {\bibinfo {author} {\bibfnamefont {T.}~\bibnamefont {Suzuki}}\ and\ \bibinfo {author} {\bibfnamefont {D.}~\bibnamefont {Rowe}},\ }\bibfield  {title} {\bibinfo {title} {The splitting of giant multipole states of deformed nuclei},\ }\href {https://doi.org/https://doi.org/10.1016/0375-9474(77)90046-X} {\bibfield  {journal} {\bibinfo  {journal} {Nucl. Phys. A}\ }\textbf {\bibinfo {volume} {289}},\ \bibinfo {pages} {461} (\bibinfo {year} {1977})}\BibitemShut {NoStop}%
\bibitem [{\citenamefont {Lo~Iudice}\ and\ \citenamefont {Palumbo}(1978)}]{Lo_Iudice_1978}%
  \BibitemOpen
  \bibfield  {author} {\bibinfo {author} {\bibfnamefont {N.}~\bibnamefont {Lo~Iudice}}\ and\ \bibinfo {author} {\bibfnamefont {F.}~\bibnamefont {Palumbo}},\ }\bibfield  {title} {\bibinfo {title} {New isovector collective modes in deformed nuclei},\ }\href {https://doi.org/10.1103/PhysRevLett.41.1532} {\bibfield  {journal} {\bibinfo  {journal} {Phys. Rev. Lett.}\ }\textbf {\bibinfo {volume} {41}},\ \bibinfo {pages} {1532} (\bibinfo {year} {1978})}\BibitemShut {NoStop}%
\bibitem [{\citenamefont {Bohle}\ \emph {et~al.}(1984)\citenamefont {Bohle}, \citenamefont {Richter}, \citenamefont {Steffen}, \citenamefont {Dieperink}, \citenamefont {{Lo Iudice}}, \citenamefont {Palumbo},\ and\ \citenamefont {Scholten}}]{Bohle1984}%
  \BibitemOpen
  \bibfield  {author} {\bibinfo {author} {\bibfnamefont {D.}~\bibnamefont {Bohle}}, \bibinfo {author} {\bibfnamefont {A.}~\bibnamefont {Richter}}, \bibinfo {author} {\bibfnamefont {W.}~\bibnamefont {Steffen}}, \bibinfo {author} {\bibfnamefont {A.}~\bibnamefont {Dieperink}}, \bibinfo {author} {\bibfnamefont {N.}~\bibnamefont {{Lo Iudice}}}, \bibinfo {author} {\bibfnamefont {F.}~\bibnamefont {Palumbo}},\ and\ \bibinfo {author} {\bibfnamefont {O.}~\bibnamefont {Scholten}},\ }\bibfield  {title} {\bibinfo {title} {New magnetic dipole excitation mode studied in the heavy deformed nucleus $^{156}\mathrm{Gd}$ by inelastic electron scattering},\ }\href {https://doi.org/https://doi.org/10.1016/0370-2693(84)91099-2} {\bibfield  {journal} {\bibinfo  {journal} {Phys. Lett. B}\ }\textbf {\bibinfo {volume} {137}},\ \bibinfo {pages} {27} (\bibinfo {year} {1984})}\BibitemShut {NoStop}%
\bibitem [{\citenamefont {Heyde}\ \emph {et~al.}(2010)\citenamefont {Heyde}, \citenamefont {von Neumann-Cosel},\ and\ \citenamefont {Richter}}]{Heyde2010}%
  \BibitemOpen
  \bibfield  {author} {\bibinfo {author} {\bibfnamefont {K.}~\bibnamefont {Heyde}}, \bibinfo {author} {\bibfnamefont {P.}~\bibnamefont {von Neumann-Cosel}},\ and\ \bibinfo {author} {\bibfnamefont {A.}~\bibnamefont {Richter}},\ }\bibfield  {title} {\bibinfo {title} {{Magnetic dipole excitations in nuclei: Elementary modes of nucleonic motion}},\ }\href {https://doi.org/10.1103/RevModPhys.82.2365} {\bibfield  {journal} {\bibinfo  {journal} {Rev. Mod. Phys.}\ }\textbf {\bibinfo {volume} {82}},\ \bibinfo {pages} {2365} (\bibinfo {year} {2010})}\BibitemShut {NoStop}%
\bibitem [{\citenamefont {Cowan}\ \emph {et~al.}(2021)\citenamefont {Cowan}, \citenamefont {Sneden}, \citenamefont {Lawler}, \citenamefont {Aprahamian}, \citenamefont {Wiescher}, \citenamefont {Langanke}, \citenamefont {Mart\'{\i}nez-Pinedo},\ and\ \citenamefont {Thielemann}}]{Cowan2021}%
  \BibitemOpen
  \bibfield  {author} {\bibinfo {author} {\bibfnamefont {J.~J.}\ \bibnamefont {Cowan}}, \bibinfo {author} {\bibfnamefont {C.}~\bibnamefont {Sneden}}, \bibinfo {author} {\bibfnamefont {J.~E.}\ \bibnamefont {Lawler}}, \bibinfo {author} {\bibfnamefont {A.}~\bibnamefont {Aprahamian}}, \bibinfo {author} {\bibfnamefont {M.}~\bibnamefont {Wiescher}}, \bibinfo {author} {\bibfnamefont {K.}~\bibnamefont {Langanke}}, \bibinfo {author} {\bibfnamefont {G.}~\bibnamefont {Mart\'{\i}nez-Pinedo}},\ and\ \bibinfo {author} {\bibfnamefont {F.-K.}\ \bibnamefont {Thielemann}},\ }\bibfield  {title} {\bibinfo {title} {{Origin of the heaviest elements: The rapid neutron-capture process}},\ }\href {https://doi.org/10.1103/RevModPhys.93.015002} {\bibfield  {journal} {\bibinfo  {journal} {Rev. Mod. Phys.}\ }\textbf {\bibinfo {volume} {93}},\ \bibinfo {pages} {015002} (\bibinfo {year} {2021})}\BibitemShut {NoStop}%
\bibitem [{\citenamefont {Eichler}\ \emph {et~al.}(2019)\citenamefont {Eichler}, \citenamefont {Sayar}, \citenamefont {Arcones},\ and\ \citenamefont {Rauscher}}]{Eichler2019}%
  \BibitemOpen
  \bibfield  {author} {\bibinfo {author} {\bibfnamefont {M.}~\bibnamefont {Eichler}}, \bibinfo {author} {\bibfnamefont {W.}~\bibnamefont {Sayar}}, \bibinfo {author} {\bibfnamefont {A.}~\bibnamefont {Arcones}},\ and\ \bibinfo {author} {\bibfnamefont {T.}~\bibnamefont {Rauscher}},\ }\bibfield  {title} {\bibinfo {title} {Probing the production of actinides under different r-process conditions},\ }\href {https://doi.org/10.3847/1538-4357/ab24cf} {\bibfield  {journal} {\bibinfo  {journal} {Astrophys. J.}\ }\textbf {\bibinfo {volume} {879}},\ \bibinfo {pages} {47} (\bibinfo {year} {2019})}\BibitemShut {NoStop}%
\bibitem [{\citenamefont {Nakatsukasa}\ \emph {et~al.}(2016)\citenamefont {Nakatsukasa}, \citenamefont {Matsuyanagi}, \citenamefont {Matsuo},\ and\ \citenamefont {Yabana}}]{Nakatsukasa2016}%
  \BibitemOpen
  \bibfield  {author} {\bibinfo {author} {\bibfnamefont {T.}~\bibnamefont {Nakatsukasa}}, \bibinfo {author} {\bibfnamefont {K.}~\bibnamefont {Matsuyanagi}}, \bibinfo {author} {\bibfnamefont {M.}~\bibnamefont {Matsuo}},\ and\ \bibinfo {author} {\bibfnamefont {K.}~\bibnamefont {Yabana}},\ }\bibfield  {title} {\bibinfo {title} {Time-dependent density-functional description of nuclear dynamics},\ }\href {https://doi.org/10.1103/RevModPhys.88.045004} {\bibfield  {journal} {\bibinfo  {journal} {Rev. Mod. Phys.}\ }\textbf {\bibinfo {volume} {88}},\ \bibinfo {pages} {045004} (\bibinfo {year} {2016})}\BibitemShut {NoStop}%
\bibitem [{\citenamefont {Nakatsukasa}\ \emph {et~al.}(2007)\citenamefont {Nakatsukasa}, \citenamefont {Inakura},\ and\ \citenamefont {Yabana}}]{Nakatsukasa2007}%
  \BibitemOpen
  \bibfield  {author} {\bibinfo {author} {\bibfnamefont {T.}~\bibnamefont {Nakatsukasa}}, \bibinfo {author} {\bibfnamefont {T.}~\bibnamefont {Inakura}},\ and\ \bibinfo {author} {\bibfnamefont {K.}~\bibnamefont {Yabana}},\ }\bibfield  {title} {\bibinfo {title} {Finite amplitude method for the solution of the random-phase approximation},\ }\href {https://doi.org/10.1103/PhysRevC.76.024318} {\bibfield  {journal} {\bibinfo  {journal} {Phys. Rev. C}\ }\textbf {\bibinfo {volume} {76}},\ \bibinfo {pages} {024318} (\bibinfo {year} {2007})}\BibitemShut {NoStop}%
\bibitem [{\citenamefont {Avogadro}\ and\ \citenamefont {Nakatsukasa}(2011)}]{Avogadro2011}%
  \BibitemOpen
  \bibfield  {author} {\bibinfo {author} {\bibfnamefont {P.}~\bibnamefont {Avogadro}}\ and\ \bibinfo {author} {\bibfnamefont {T.}~\bibnamefont {Nakatsukasa}},\ }\bibfield  {title} {\bibinfo {title} {Finite amplitude method for the quasiparticle random-phase approximation},\ }\href {https://doi.org/10.1103/PhysRevC.84.014314} {\bibfield  {journal} {\bibinfo  {journal} {Phys. Rev. C}\ }\textbf {\bibinfo {volume} {84}},\ \bibinfo {pages} {014314} (\bibinfo {year} {2011})}\BibitemShut {NoStop}%
\bibitem [{\citenamefont {Oishi}\ \emph {et~al.}(2016)\citenamefont {Oishi}, \citenamefont {Kortelainen},\ and\ \citenamefont {Hinohara}}]{Oishi2016}%
  \BibitemOpen
  \bibfield  {author} {\bibinfo {author} {\bibfnamefont {T.}~\bibnamefont {Oishi}}, \bibinfo {author} {\bibfnamefont {M.}~\bibnamefont {Kortelainen}},\ and\ \bibinfo {author} {\bibfnamefont {N.}~\bibnamefont {Hinohara}},\ }\bibfield  {title} {\bibinfo {title} {Finite amplitude method applied to the giant dipole resonance in heavy rare-earth nuclei},\ }\href {https://doi.org/10.1103/PhysRevC.93.034329} {\bibfield  {journal} {\bibinfo  {journal} {Phys. Rev. C}\ }\textbf {\bibinfo {volume} {93}},\ \bibinfo {pages} {034329} (\bibinfo {year} {2016})}\BibitemShut {NoStop}%
\bibitem [{\citenamefont {Li}\ \emph {et~al.}(2024)\citenamefont {Li}, \citenamefont {Schunck},\ and\ \citenamefont {Grosskopf}}]{Tong2024}%
  \BibitemOpen
  \bibfield  {author} {\bibinfo {author} {\bibfnamefont {T.}~\bibnamefont {Li}}, \bibinfo {author} {\bibfnamefont {N.}~\bibnamefont {Schunck}},\ and\ \bibinfo {author} {\bibfnamefont {M.}~\bibnamefont {Grosskopf}},\ }\bibfield  {title} {\bibinfo {title} {Multipole responses in fissioning nuclei and their uncertainties},\ }\href {https://doi.org/10.1103/PhysRevC.110.034317} {\bibfield  {journal} {\bibinfo  {journal} {Phys. Rev. C}\ }\textbf {\bibinfo {volume} {110}},\ \bibinfo {pages} {034317} (\bibinfo {year} {2024})}\BibitemShut {NoStop}%
\bibitem [{\citenamefont {Washiyama}\ \emph {et~al.}(2021)\citenamefont {Washiyama}, \citenamefont {Hinohara},\ and\ \citenamefont {Nakatsukasa}}]{Washiyama2021}%
  \BibitemOpen
  \bibfield  {author} {\bibinfo {author} {\bibfnamefont {K.}~\bibnamefont {Washiyama}}, \bibinfo {author} {\bibfnamefont {N.}~\bibnamefont {Hinohara}},\ and\ \bibinfo {author} {\bibfnamefont {T.}~\bibnamefont {Nakatsukasa}},\ }\bibfield  {title} {\bibinfo {title} {Finite-amplitude method for collective inertia in spontaneous fission},\ }\href {https://doi.org/10.1103/PhysRevC.103.014306} {\bibfield  {journal} {\bibinfo  {journal} {Phys. Rev. C}\ }\textbf {\bibinfo {volume} {103}},\ \bibinfo {pages} {014306} (\bibinfo {year} {2021})}\BibitemShut {NoStop}%
\bibitem [{\citenamefont {Hinohara}(2015)}]{Hinohara2015}%
  \BibitemOpen
  \bibfield  {author} {\bibinfo {author} {\bibfnamefont {N.}~\bibnamefont {Hinohara}},\ }\bibfield  {title} {\bibinfo {title} {Collective inertia of the {Nambu-Goldstone} mode from linear response theory},\ }\href {https://doi.org/10.1103/PhysRevC.92.034321} {\bibfield  {journal} {\bibinfo  {journal} {Phys. Rev. C}\ }\textbf {\bibinfo {volume} {92}},\ \bibinfo {pages} {034321} (\bibinfo {year} {2015})}\BibitemShut {NoStop}%
\bibitem [{\citenamefont {Hinohara}\ and\ \citenamefont {Nazarewicz}(2016)}]{Hinohara2016}%
  \BibitemOpen
  \bibfield  {author} {\bibinfo {author} {\bibfnamefont {N.}~\bibnamefont {Hinohara}}\ and\ \bibinfo {author} {\bibfnamefont {W.}~\bibnamefont {Nazarewicz}},\ }\bibfield  {title} {\bibinfo {title} {{Pairing Nambu-Goldstone Modes within Nuclear Density Functional Theory}},\ }\href {https://doi.org/10.1103/PhysRevLett.116.152502} {\bibfield  {journal} {\bibinfo  {journal} {Phys. Rev. Lett.}\ }\textbf {\bibinfo {volume} {116}},\ \bibinfo {pages} {152502} (\bibinfo {year} {2016})}\BibitemShut {NoStop}%
\bibitem [{\citenamefont {Petr\'{\i}k}\ and\ \citenamefont {Kortelainen}(2018)}]{Petrik2018}%
  \BibitemOpen
  \bibfield  {author} {\bibinfo {author} {\bibfnamefont {K.}~\bibnamefont {Petr\'{\i}k}}\ and\ \bibinfo {author} {\bibfnamefont {M.}~\bibnamefont {Kortelainen}},\ }\bibfield  {title} {\bibinfo {title} {{Thouless-Valatin} rotational moment of inertia from linear response theory},\ }\href {https://doi.org/10.1103/PhysRevC.97.034321} {\bibfield  {journal} {\bibinfo  {journal} {Phys. Rev. C}\ }\textbf {\bibinfo {volume} {97}},\ \bibinfo {pages} {034321} (\bibinfo {year} {2018})}\BibitemShut {NoStop}%
\bibitem [{\citenamefont {Shafer}\ \emph {et~al.}(2016)\citenamefont {Shafer}, \citenamefont {Engel}, \citenamefont {Fr\"ohlich}, \citenamefont {McLaughlin}, \citenamefont {Mumpower},\ and\ \citenamefont {Surman}}]{Shafer2016}%
  \BibitemOpen
  \bibfield  {author} {\bibinfo {author} {\bibfnamefont {T.}~\bibnamefont {Shafer}}, \bibinfo {author} {\bibfnamefont {J.}~\bibnamefont {Engel}}, \bibinfo {author} {\bibfnamefont {C.}~\bibnamefont {Fr\"ohlich}}, \bibinfo {author} {\bibfnamefont {G.~C.}\ \bibnamefont {McLaughlin}}, \bibinfo {author} {\bibfnamefont {M.}~\bibnamefont {Mumpower}},\ and\ \bibinfo {author} {\bibfnamefont {R.}~\bibnamefont {Surman}},\ }\bibfield  {title} {\bibinfo {title} {{$\ensuremath{\beta}$ decay of deformed $r$-process nuclei near $A=80$ and $A=160$, including odd-$A$ and odd-odd nuclei, with the Skyrme finite-amplitude method}},\ }\href {https://doi.org/10.1103/PhysRevC.94.055802} {\bibfield  {journal} {\bibinfo  {journal} {Phys. Rev. C}\ }\textbf {\bibinfo {volume} {94}},\ \bibinfo {pages} {055802} (\bibinfo {year} {2016})}\BibitemShut {NoStop}%
\bibitem [{\citenamefont {Hinohara}\ and\ \citenamefont {Engel}(2022)}]{Hinohara2022}%
  \BibitemOpen
  \bibfield  {author} {\bibinfo {author} {\bibfnamefont {N.}~\bibnamefont {Hinohara}}\ and\ \bibinfo {author} {\bibfnamefont {J.}~\bibnamefont {Engel}},\ }\bibfield  {title} {\bibinfo {title} {Global calculation of two-neutrino double-$\ensuremath{\beta}$ decay within the finite amplitude method in nuclear density functional theory},\ }\href {https://doi.org/10.1103/PhysRevC.105.044314} {\bibfield  {journal} {\bibinfo  {journal} {Phys. Rev. C}\ }\textbf {\bibinfo {volume} {105}},\ \bibinfo {pages} {044314} (\bibinfo {year} {2022})}\BibitemShut {NoStop}%
\bibitem [{\citenamefont {Zhao}(2024)}]{Zhao2024}%
  \BibitemOpen
  \bibfield  {author} {\bibinfo {author} {\bibfnamefont {J.}~\bibnamefont {Zhao}},\ }\bibfield  {title} {\bibinfo {title} {{Multipole modes of excitation in tetrahedrally deformed neutron-rich Zr isotopes}},\ }\href {https://doi.org/10.1103/PhysRevC.110.L011301} {\bibfield  {journal} {\bibinfo  {journal} {Phys. Rev. C}\ }\textbf {\bibinfo {volume} {110}},\ \bibinfo {pages} {L011301} (\bibinfo {year} {2024})}\BibitemShut {NoStop}%
\bibitem [{\citenamefont {Zamora}\ \emph {et~al.}(2016)\citenamefont {Zamora} \emph {et~al.}}]{Zamora2016}%
  \BibitemOpen
  \bibfield  {author} {\bibinfo {author} {\bibfnamefont {J.}~\bibnamefont {Zamora}} \emph {et~al.},\ }\bibfield  {title} {\bibinfo {title} {First measurement of isoscalar giant resonances in a stored-beam experiment},\ }\href {https://doi.org/https://doi.org/10.1016/j.physletb.2016.10.015} {\bibfield  {journal} {\bibinfo  {journal} {Phys. Lett. B}\ }\textbf {\bibinfo {volume} {763}},\ \bibinfo {pages} {16} (\bibinfo {year} {2016})}\BibitemShut {NoStop}%
\bibitem [{\citenamefont {Reifarth}\ and\ \citenamefont {Litvinov}(2014)}]{Reinhart2014}%
  \BibitemOpen
  \bibfield  {author} {\bibinfo {author} {\bibfnamefont {R.}~\bibnamefont {Reifarth}}\ and\ \bibinfo {author} {\bibfnamefont {Y.~A.}\ \bibnamefont {Litvinov}},\ }\bibfield  {title} {\bibinfo {title} {Measurements of neutron-induced reactions in inverse kinematics},\ }\href {https://doi.org/10.1103/PhysRevSTAB.17.014701} {\bibfield  {journal} {\bibinfo  {journal} {Phys. Rev. ST Accel. Beams}\ }\textbf {\bibinfo {volume} {17}},\ \bibinfo {pages} {014701} (\bibinfo {year} {2014})}\BibitemShut {NoStop}%
\bibitem [{\citenamefont {Reifarth}\ \emph {et~al.}(2017)\citenamefont {Reifarth}, \citenamefont {G\"obel}, \citenamefont {Heftrich}, \citenamefont {Weigand}, \citenamefont {Jurado}, \citenamefont {K\"appeler},\ and\ \citenamefont {Litvinov}}]{Reifarth2017}%
  \BibitemOpen
  \bibfield  {author} {\bibinfo {author} {\bibfnamefont {R.}~\bibnamefont {Reifarth}}, \bibinfo {author} {\bibfnamefont {K.}~\bibnamefont {G\"obel}}, \bibinfo {author} {\bibfnamefont {T.}~\bibnamefont {Heftrich}}, \bibinfo {author} {\bibfnamefont {M.}~\bibnamefont {Weigand}}, \bibinfo {author} {\bibfnamefont {B.}~\bibnamefont {Jurado}}, \bibinfo {author} {\bibfnamefont {F.}~\bibnamefont {K\"appeler}},\ and\ \bibinfo {author} {\bibfnamefont {Y.~A.}\ \bibnamefont {Litvinov}},\ }\bibfield  {title} {\bibinfo {title} {Spallation-based neutron target for direct studies of neutron-induced reactions in inverse kinematics},\ }\href {https://doi.org/10.1103/PhysRevAccelBeams.20.044701} {\bibfield  {journal} {\bibinfo  {journal} {Phys. Rev. Accel. Beams}\ }\textbf {\bibinfo {volume} {20}},\ \bibinfo {pages} {044701} (\bibinfo {year} {2017})}\BibitemShut {NoStop}%
\bibitem [{\citenamefont {Cooper}\ \emph {et~al.}(2024)\citenamefont {Cooper}, \citenamefont {Mosby}, \citenamefont {Reifarth}, \citenamefont {Couture}, \citenamefont {Bennett}, \citenamefont {Gibson}, \citenamefont {Gorelov}, \citenamefont {Keith}, \citenamefont {Lovell}, \citenamefont {Misch},\ and\ \citenamefont {Mumpower}}]{Cooper2024}%
  \BibitemOpen
  \bibfield  {author} {\bibinfo {author} {\bibfnamefont {A.~L.}\ \bibnamefont {Cooper}}, \bibinfo {author} {\bibfnamefont {S.}~\bibnamefont {Mosby}}, \bibinfo {author} {\bibfnamefont {R.}~\bibnamefont {Reifarth}}, \bibinfo {author} {\bibfnamefont {A.}~\bibnamefont {Couture}}, \bibinfo {author} {\bibfnamefont {E.}~\bibnamefont {Bennett}}, \bibinfo {author} {\bibfnamefont {N.}~\bibnamefont {Gibson}}, \bibinfo {author} {\bibfnamefont {D.}~\bibnamefont {Gorelov}}, \bibinfo {author} {\bibfnamefont {C.}~\bibnamefont {Keith}}, \bibinfo {author} {\bibfnamefont {A.}~\bibnamefont {Lovell}}, \bibinfo {author} {\bibfnamefont {G.}~\bibnamefont {Misch}},\ and\ \bibinfo {author} {\bibfnamefont {M.}~\bibnamefont {Mumpower}},\ }\bibfield  {title} {\bibinfo {title} {A high-intensity, low-energy heavy ion source for a neutron target proof-of-principle experiment at lansce},\ }\href {https://doi.org/10.1088/1742-6596/2743/1/012091} {\bibfield  {journal} {\bibinfo  {journal} {J. Phys.:Conf. Ser.}\ }\textbf {\bibinfo {volume} {2743}},\ \bibinfo {pages} {012091} (\bibinfo {year} {2024})}\BibitemShut {NoStop}%
\bibitem [{\citenamefont {Stoitsov}\ \emph {et~al.}(2013)\citenamefont {Stoitsov}, \citenamefont {Schunck}, \citenamefont {Kortelainen}, \citenamefont {Michel}, \citenamefont {Nam}, \citenamefont {Olsen}, \citenamefont {Sarich},\ and\ \citenamefont {Wild}}]{hfbtho_v2_00}%
  \BibitemOpen
  \bibfield  {author} {\bibinfo {author} {\bibfnamefont {M.}~\bibnamefont {Stoitsov}}, \bibinfo {author} {\bibfnamefont {N.}~\bibnamefont {Schunck}}, \bibinfo {author} {\bibfnamefont {M.}~\bibnamefont {Kortelainen}}, \bibinfo {author} {\bibfnamefont {N.}~\bibnamefont {Michel}}, \bibinfo {author} {\bibfnamefont {H.}~\bibnamefont {Nam}}, \bibinfo {author} {\bibfnamefont {E.}~\bibnamefont {Olsen}}, \bibinfo {author} {\bibfnamefont {J.}~\bibnamefont {Sarich}},\ and\ \bibinfo {author} {\bibfnamefont {S.}~\bibnamefont {Wild}},\ }\bibfield  {title} {\bibinfo {title} {{Axially deformed solution of the Skyrme-Hartree–Fock–Bogoliubov equations using the transformed harmonic oscillator basis (II) hfbtho v2.00d: A new version of the program}},\ }\href {https://doi.org/https://doi.org/10.1016/j.cpc.2013.01.013} {\bibfield  {journal} {\bibinfo  {journal} {Comput. Phys. Commun.}\ }\textbf {\bibinfo {volume} {184}},\ \bibinfo {pages} {1592} (\bibinfo {year} {2013})}\BibitemShut {NoStop}%
\bibitem [{\citenamefont {Stoitsov}\ \emph {et~al.}(2011)\citenamefont {Stoitsov}, \citenamefont {Kortelainen}, \citenamefont {Nakatsukasa}, \citenamefont {Losa},\ and\ \citenamefont {Nazarewicz}}]{Stoitsov2011}%
  \BibitemOpen
  \bibfield  {author} {\bibinfo {author} {\bibfnamefont {M.}~\bibnamefont {Stoitsov}}, \bibinfo {author} {\bibfnamefont {M.}~\bibnamefont {Kortelainen}}, \bibinfo {author} {\bibfnamefont {T.}~\bibnamefont {Nakatsukasa}}, \bibinfo {author} {\bibfnamefont {C.}~\bibnamefont {Losa}},\ and\ \bibinfo {author} {\bibfnamefont {W.}~\bibnamefont {Nazarewicz}},\ }\bibfield  {title} {\bibinfo {title} {Monopole strength function of deformed superfluid nuclei},\ }\href {https://doi.org/10.1103/PhysRevC.84.041305} {\bibfield  {journal} {\bibinfo  {journal} {Phys. Rev. C}\ }\textbf {\bibinfo {volume} {84}},\ \bibinfo {pages} {041305(R)} (\bibinfo {year} {2011})}\BibitemShut {NoStop}%
\bibitem [{\citenamefont {Kortelainen}\ \emph {et~al.}(2015)\citenamefont {Kortelainen}, \citenamefont {Hinohara},\ and\ \citenamefont {Nazarewicz}}]{Kortelainen2015}%
  \BibitemOpen
  \bibfield  {author} {\bibinfo {author} {\bibfnamefont {M.}~\bibnamefont {Kortelainen}}, \bibinfo {author} {\bibfnamefont {N.}~\bibnamefont {Hinohara}},\ and\ \bibinfo {author} {\bibfnamefont {W.}~\bibnamefont {Nazarewicz}},\ }\bibfield  {title} {\bibinfo {title} {Multipole modes in deformed nuclei within the finite amplitude method},\ }\href {https://doi.org/10.1103/PhysRevC.92.051302} {\bibfield  {journal} {\bibinfo  {journal} {Phys. Rev. C}\ }\textbf {\bibinfo {volume} {92}},\ \bibinfo {pages} {051302(R)} (\bibinfo {year} {2015})}\BibitemShut {NoStop}%
\bibitem [{Sup()}]{Supplemental_Material}%
  \BibitemOpen
  \href@noop {} {}\bibinfo {note} {See Supplemental Material for additional figures and further analyses.}\BibitemShut {Stop}%
\bibitem [{\citenamefont {Scamps}\ and\ \citenamefont {Lacroix}(2013)}]{Scamps2013}%
  \BibitemOpen
  \bibfield  {author} {\bibinfo {author} {\bibfnamefont {G.}~\bibnamefont {Scamps}}\ and\ \bibinfo {author} {\bibfnamefont {D.}~\bibnamefont {Lacroix}},\ }\bibfield  {title} {\bibinfo {title} {Systematics of isovector and isoscalar giant quadrupole resonances in normal and superfluid spherical nuclei},\ }\href {https://doi.org/10.1103/PhysRevC.88.044310} {\bibfield  {journal} {\bibinfo  {journal} {Phys. Rev. C}\ }\textbf {\bibinfo {volume} {88}},\ \bibinfo {pages} {044310} (\bibinfo {year} {2013})}\BibitemShut {NoStop}%
\bibitem [{\citenamefont {Younes}\ and\ \citenamefont {Gogny}(2009)}]{Younes2009}%
  \BibitemOpen
  \bibfield  {author} {\bibinfo {author} {\bibfnamefont {W.}~\bibnamefont {Younes}}\ and\ \bibinfo {author} {\bibfnamefont {D.}~\bibnamefont {Gogny}},\ }\bibfield  {title} {\bibinfo {title} {{Microscopic calculation of $^{240}\mathrm{Pu}$ scission with a finite-range effective force}},\ }\href {https://doi.org/10.1103/PhysRevC.80.054313} {\bibfield  {journal} {\bibinfo  {journal} {Phys. Rev. C}\ }\textbf {\bibinfo {volume} {80}},\ \bibinfo {pages} {054313} (\bibinfo {year} {2009})}\BibitemShut {NoStop}%
\bibitem [{\citenamefont {Kortelainen}(2020)}]{Kortelainen2020}%
  \BibitemOpen
  \bibfield  {author} {\bibinfo {author} {\bibfnamefont {M.}~\bibnamefont {Kortelainen}},\ }\bibfield  {title} {\bibinfo {title} {{Thouless-Valatin moment of inertia and removal of the spurious mode in the linear response theory}},\ }\href {https://doi.org/10.1088/1742-6596/1643/1/012142} {\bibfield  {journal} {\bibinfo  {journal} {J. Phys.:Conf. Ser.}\ }\textbf {\bibinfo {volume} {1643}},\ \bibinfo {pages} {012142} (\bibinfo {year} {2020})}\BibitemShut {NoStop}%
\bibitem [{\citenamefont {Bartel}\ \emph {et~al.}(1982)\citenamefont {Bartel}, \citenamefont {Quentin}, \citenamefont {Brack}, \citenamefont {Guet},\ and\ \citenamefont {Håkansson}}]{Bartel1982}%
  \BibitemOpen
  \bibfield  {author} {\bibinfo {author} {\bibfnamefont {J.}~\bibnamefont {Bartel}}, \bibinfo {author} {\bibfnamefont {P.}~\bibnamefont {Quentin}}, \bibinfo {author} {\bibfnamefont {M.}~\bibnamefont {Brack}}, \bibinfo {author} {\bibfnamefont {C.}~\bibnamefont {Guet}},\ and\ \bibinfo {author} {\bibfnamefont {H.-B.}\ \bibnamefont {Håkansson}},\ }\bibfield  {title} {\bibinfo {title} {{Towards a better parametrisation of Skyrme-like effective forces: A critical study of the SkM force}},\ }\href {https://doi.org/https://doi.org/10.1016/0375-9474(82)90403-1} {\bibfield  {journal} {\bibinfo  {journal} {Nucl. Phys. A}\ }\textbf {\bibinfo {volume} {386}},\ \bibinfo {pages} {79} (\bibinfo {year} {1982})}\BibitemShut {NoStop}%
\bibitem [{\citenamefont {Chabanat}\ \emph {et~al.}(1998)\citenamefont {Chabanat}, \citenamefont {Bonche}, \citenamefont {Haensel}, \citenamefont {Meyer},\ and\ \citenamefont {Schaeffer}}]{Chabanat1998}%
  \BibitemOpen
  \bibfield  {author} {\bibinfo {author} {\bibfnamefont {E.}~\bibnamefont {Chabanat}}, \bibinfo {author} {\bibfnamefont {P.}~\bibnamefont {Bonche}}, \bibinfo {author} {\bibfnamefont {P.}~\bibnamefont {Haensel}}, \bibinfo {author} {\bibfnamefont {J.}~\bibnamefont {Meyer}},\ and\ \bibinfo {author} {\bibfnamefont {R.}~\bibnamefont {Schaeffer}},\ }\bibfield  {title} {\bibinfo {title} {{A Skyrme parametrization from subnuclear to neutron star densities Part II. Nuclei far from stabilities}},\ }\href {https://doi.org/10.1016/S0375-9474(98)00180-8} {\bibfield  {journal} {\bibinfo  {journal} {Nucl. Phys. A}\ }\textbf {\bibinfo {volume} {635}},\ \bibinfo {pages} {231} (\bibinfo {year} {1998})}\BibitemShut {NoStop}%
\bibitem [{\citenamefont {Kortelainen}\ \emph {et~al.}(2012)\citenamefont {Kortelainen}, \citenamefont {McDonnell}, \citenamefont {Nazarewicz}, \citenamefont {Reinhard}, \citenamefont {Sarich}, \citenamefont {Schunck}, \citenamefont {Stoitsov},\ and\ \citenamefont {Wild}}]{Kortelainen2012}%
  \BibitemOpen
  \bibfield  {author} {\bibinfo {author} {\bibfnamefont {M.}~\bibnamefont {Kortelainen}}, \bibinfo {author} {\bibfnamefont {J.}~\bibnamefont {McDonnell}}, \bibinfo {author} {\bibfnamefont {W.}~\bibnamefont {Nazarewicz}}, \bibinfo {author} {\bibfnamefont {P.-G.}\ \bibnamefont {Reinhard}}, \bibinfo {author} {\bibfnamefont {J.}~\bibnamefont {Sarich}}, \bibinfo {author} {\bibfnamefont {N.}~\bibnamefont {Schunck}}, \bibinfo {author} {\bibfnamefont {M.~V.}\ \bibnamefont {Stoitsov}},\ and\ \bibinfo {author} {\bibfnamefont {S.~M.}\ \bibnamefont {Wild}},\ }\bibfield  {title} {\bibinfo {title} {{Nuclear energy density optimization: Large deformations}},\ }\href {https://doi.org/10.1103/PhysRevC.85.024304} {\bibfield  {journal} {\bibinfo  {journal} {Phys. Rev. C}\ }\textbf {\bibinfo {volume} {85}},\ \bibinfo {pages} {024304} (\bibinfo {year} {2012})}\BibitemShut {NoStop}%
\bibitem [{\citenamefont {Schunck}\ \emph {et~al.}(2014)\citenamefont {Schunck}, \citenamefont {Duke}, \citenamefont {Carr},\ and\ \citenamefont {Knoll}}]{Schunck2014}%
  \BibitemOpen
  \bibfield  {author} {\bibinfo {author} {\bibfnamefont {N.}~\bibnamefont {Schunck}}, \bibinfo {author} {\bibfnamefont {D.}~\bibnamefont {Duke}}, \bibinfo {author} {\bibfnamefont {H.}~\bibnamefont {Carr}},\ and\ \bibinfo {author} {\bibfnamefont {A.}~\bibnamefont {Knoll}},\ }\bibfield  {title} {\bibinfo {title} {{Description of induced nuclear fission with Skyrme energy functionals: Static potential energy surfaces and fission fragment properties}},\ }\href {https://doi.org/10.1103/PhysRevC.90.054305} {\bibfield  {journal} {\bibinfo  {journal} {Phys. Rev. C}\ }\textbf {\bibinfo {volume} {90}},\ \bibinfo {pages} {054305} (\bibinfo {year} {2014})}\BibitemShut {NoStop}%
\bibitem [{\citenamefont {Bonnard}\ \emph {et~al.}(2023)\citenamefont {Bonnard}, \citenamefont {Dobaczewski}, \citenamefont {Danneaux},\ and\ \citenamefont {Kortelainen}}]{Bonnard2023}%
  \BibitemOpen
  \bibfield  {author} {\bibinfo {author} {\bibfnamefont {J.}~\bibnamefont {Bonnard}}, \bibinfo {author} {\bibfnamefont {J.}~\bibnamefont {Dobaczewski}}, \bibinfo {author} {\bibfnamefont {G.}~\bibnamefont {Danneaux}},\ and\ \bibinfo {author} {\bibfnamefont {M.}~\bibnamefont {Kortelainen}},\ }\bibfield  {title} {\bibinfo {title} {Nuclear {DFT} electromagnetic moments in heavy deformed open-shell odd nuclei},\ }\href {https://doi.org/https://doi.org/10.1016/j.physletb.2023.138014} {\bibfield  {journal} {\bibinfo  {journal} {Phys. Lett. B}\ }\textbf {\bibinfo {volume} {843}},\ \bibinfo {pages} {138014} (\bibinfo {year} {2023})}\BibitemShut {NoStop}%
\bibitem [{\citenamefont {Bender}\ \emph {et~al.}(2002)\citenamefont {Bender}, \citenamefont {Dobaczewski}, \citenamefont {Engel},\ and\ \citenamefont {Nazarewicz}}]{Bender2002}%
  \BibitemOpen
  \bibfield  {author} {\bibinfo {author} {\bibfnamefont {M.}~\bibnamefont {Bender}}, \bibinfo {author} {\bibfnamefont {J.}~\bibnamefont {Dobaczewski}}, \bibinfo {author} {\bibfnamefont {J.}~\bibnamefont {Engel}},\ and\ \bibinfo {author} {\bibfnamefont {W.}~\bibnamefont {Nazarewicz}},\ }\bibfield  {title} {\bibinfo {title} {{Gamow-Teller strength and the spin-isospin coupling constants of the Skyrme energy functional}},\ }\href {https://doi.org/10.1103/PhysRevC.65.054322} {\bibfield  {journal} {\bibinfo  {journal} {Phys. Rev. C}\ }\textbf {\bibinfo {volume} {65}},\ \bibinfo {pages} {054322} (\bibinfo {year} {2002})}\BibitemShut {NoStop}%
\bibitem [{\citenamefont {Sassarini}\ \emph {et~al.}(2022)\citenamefont {Sassarini}, \citenamefont {Dobaczewski}, \citenamefont {Bonnard},\ and\ \citenamefont {Ruiz}}]{Sassarini_2022}%
  \BibitemOpen
  \bibfield  {author} {\bibinfo {author} {\bibfnamefont {P.~L.}\ \bibnamefont {Sassarini}}, \bibinfo {author} {\bibfnamefont {J.}~\bibnamefont {Dobaczewski}}, \bibinfo {author} {\bibfnamefont {J.}~\bibnamefont {Bonnard}},\ and\ \bibinfo {author} {\bibfnamefont {R.~F.~G.}\ \bibnamefont {Ruiz}},\ }\bibfield  {title} {\bibinfo {title} {{Nuclear DFT analysis of electromagnetic moments in odd near doubly magic nuclei}},\ }\href {https://doi.org/10.1088/1361-6471/ac900a} {\bibfield  {journal} {\bibinfo  {journal} {J. Phys. G:Nucl. Part. Phys.}\ }\textbf {\bibinfo {volume} {49}},\ \bibinfo {pages} {11LT01} (\bibinfo {year} {2022})}\BibitemShut {NoStop}%
\bibitem [{\citenamefont {Nesterenko}\ \emph {et~al.}(2007)\citenamefont {Nesterenko}, \citenamefont {Kleinig}, \citenamefont {Kvasil}, \citenamefont {Vesely},\ and\ \citenamefont {Reinhard}}]{Nesterenko2007}%
  \BibitemOpen
  \bibfield  {author} {\bibinfo {author} {\bibfnamefont {V.~O.}\ \bibnamefont {Nesterenko}}, \bibinfo {author} {\bibfnamefont {W.}~\bibnamefont {Kleinig}}, \bibinfo {author} {\bibfnamefont {J.}~\bibnamefont {Kvasil}}, \bibinfo {author} {\bibfnamefont {P.}~\bibnamefont {Vesely}},\ and\ \bibinfo {author} {\bibfnamefont {P.-G.}\ \bibnamefont {Reinhard}},\ }\bibfield  {title} {\bibinfo {title} {{Giant dipole resonance in deformed nuclei: Dependence on Skyrme forces}},\ }\href {https://doi.org/10.1142/S0218301307006071} {\bibfield  {journal} {\bibinfo  {journal} {Int. J. Mod. Phys. E}\ }\textbf {\bibinfo {volume} {16}},\ \bibinfo {pages} {624} (\bibinfo {year} {2007})}\BibitemShut {NoStop}%
\bibitem [{\citenamefont {Richter}(1995)}]{Richter1995}%
  \BibitemOpen
  \bibfield  {author} {\bibinfo {author} {\bibfnamefont {A.}~\bibnamefont {Richter}},\ }\bibfield  {title} {\bibinfo {title} {Probing the nuclear magnetic dipole response with electrons, photons and hadrons},\ }\href {https://doi.org/https://doi.org/10.1016/0146-6410(95)00022-B} {\bibfield  {journal} {\bibinfo  {journal} {Prog. Part. Nucl. Phys.}\ }\textbf {\bibinfo {volume} {34}},\ \bibinfo {pages} {261} (\bibinfo {year} {1995})}\BibitemShut {NoStop}%
\bibitem [{\citenamefont {Vesely}\ \emph {et~al.}(2009)\citenamefont {Vesely}, \citenamefont {Kvasil}, \citenamefont {Nesterenko}, \citenamefont {Kleinig}, \citenamefont {Reinhard},\ and\ \citenamefont {Ponomarev}}]{Vesely2009}%
  \BibitemOpen
  \bibfield  {author} {\bibinfo {author} {\bibfnamefont {P.}~\bibnamefont {Vesely}}, \bibinfo {author} {\bibfnamefont {J.}~\bibnamefont {Kvasil}}, \bibinfo {author} {\bibfnamefont {V.~O.}\ \bibnamefont {Nesterenko}}, \bibinfo {author} {\bibfnamefont {W.}~\bibnamefont {Kleinig}}, \bibinfo {author} {\bibfnamefont {P.~G.}\ \bibnamefont {Reinhard}},\ and\ \bibinfo {author} {\bibfnamefont {V.~Y.}\ \bibnamefont {Ponomarev}},\ }\bibfield  {title} {\bibinfo {title} {{Skyrme random-phase-approximation description of spin-flip $M1$ giant resonance}},\ }\href {https://doi.org/10.1103/PhysRevC.80.031302} {\bibfield  {journal} {\bibinfo  {journal} {Phys. Rev. C}\ }\textbf {\bibinfo {volume} {80}},\ \bibinfo {pages} {031302} (\bibinfo {year} {2009})}\BibitemShut {NoStop}%
\bibitem [{\citenamefont {Ziegler}\ \emph {et~al.}(1990)\citenamefont {Ziegler}, \citenamefont {Rangacharyulu}, \citenamefont {Richter},\ and\ \citenamefont {Spieler}}]{Ziegler1990}%
  \BibitemOpen
  \bibfield  {author} {\bibinfo {author} {\bibfnamefont {W.}~\bibnamefont {Ziegler}}, \bibinfo {author} {\bibfnamefont {C.}~\bibnamefont {Rangacharyulu}}, \bibinfo {author} {\bibfnamefont {A.}~\bibnamefont {Richter}},\ and\ \bibinfo {author} {\bibfnamefont {C.}~\bibnamefont {Spieler}},\ }\bibfield  {title} {\bibinfo {title} {{Orbital magnetic dipole strength in $^{148,150,152,154}\mathrm{Sm}$ and nuclear deformation}},\ }\href {https://doi.org/10.1103/PhysRevLett.65.2515} {\bibfield  {journal} {\bibinfo  {journal} {Phys. Rev. Lett.}\ }\textbf {\bibinfo {volume} {65}},\ \bibinfo {pages} {2515} (\bibinfo {year} {1990})}\BibitemShut {NoStop}%
\bibitem [{\citenamefont {Dobaczewski}\ \emph {et~al.}(2026)\citenamefont {Dobaczewski}, \citenamefont {Stuchbery}, \citenamefont {Danneaux}, \citenamefont {Nagpal}, \citenamefont {Sassarini},\ and\ \citenamefont {Wibowo}}]{Dobaczewski2026}%
  \BibitemOpen
  \bibfield  {author} {\bibinfo {author} {\bibfnamefont {J.}~\bibnamefont {Dobaczewski}}, \bibinfo {author} {\bibfnamefont {A.~E.}\ \bibnamefont {Stuchbery}}, \bibinfo {author} {\bibfnamefont {G.}~\bibnamefont {Danneaux}}, \bibinfo {author} {\bibfnamefont {A.}~\bibnamefont {Nagpal}}, \bibinfo {author} {\bibfnamefont {P.~L.}\ \bibnamefont {Sassarini}},\ and\ \bibinfo {author} {\bibfnamefont {H.}~\bibnamefont {Wibowo}},\ }\bibfield  {title} {\bibinfo {title} {{Electromagnetic moments of ground and excited states calculated in heavy odd-$N$ open-shell nuclei}},\ }\href {https://doi.org/10.1103/4q99-rv67} {\bibfield  {journal} {\bibinfo  {journal} {Phys. Rev. C}\ }\textbf {\bibinfo {volume} {113}},\ \bibinfo {pages} {024306} (\bibinfo {year} {2026})}\BibitemShut {NoStop}%
\bibitem [{\citenamefont {Lo~Iudice}(1997)}]{Lo_Iudice1997}%
  \BibitemOpen
  \bibfield  {author} {\bibinfo {author} {\bibfnamefont {N.}~\bibnamefont {Lo~Iudice}},\ }\bibfield  {title} {\bibinfo {title} {Magnetic dipole excitations in deformed nuclei},\ }\href {https://doi.org/10.1134/1.953055} {\bibfield  {journal} {\bibinfo  {journal} {Phys. Part. Nucl.}\ }\textbf {\bibinfo {volume} {28}},\ \bibinfo {pages} {556} (\bibinfo {year} {1997})}\BibitemShut {NoStop}%
\bibitem [{\citenamefont {Kurath}(1963)}]{Kurath1963}%
  \BibitemOpen
  \bibfield  {author} {\bibinfo {author} {\bibfnamefont {D.}~\bibnamefont {Kurath}},\ }\bibfield  {title} {\bibinfo {title} {{Strong $M1$ Transitions in Light Nuclei}},\ }\href {https://doi.org/10.1103/PhysRev.130.1525} {\bibfield  {journal} {\bibinfo  {journal} {Phys. Rev.}\ }\textbf {\bibinfo {volume} {130}},\ \bibinfo {pages} {1525} (\bibinfo {year} {1963})}\BibitemShut {NoStop}%
\bibitem [{\citenamefont {Lo~Iudice}(2000)}]{Lo_Iudice2000}%
  \BibitemOpen
  \bibfield  {author} {\bibinfo {author} {\bibfnamefont {N.}~\bibnamefont {Lo~Iudice}},\ }\bibfield  {title} {\bibinfo {title} {Collective excitations in deformed nuclei},\ }\href {https://doi.org/10.1007/BF03548889} {\bibfield  {journal} {\bibinfo  {journal} {Riv.Nuovo Cimento}\ }\textbf {\bibinfo {volume} {23}},\ \bibinfo {pages} {1} (\bibinfo {year} {2000})}\BibitemShut {NoStop}%
\bibitem [{\citenamefont {Reinhard}(1999)}]{Reinhard1999}%
  \BibitemOpen
  \bibfield  {author} {\bibinfo {author} {\bibfnamefont {P.-G.}\ \bibnamefont {Reinhard}},\ }\bibfield  {title} {\bibinfo {title} {Skyrme forces and giant resonances in exotic nuclei},\ }\href {https://doi.org/https://doi.org/10.1016/S0375-9474(99)00076-7} {\bibfield  {journal} {\bibinfo  {journal} {Nucl. Phys. A}\ }\textbf {\bibinfo {volume} {649}},\ \bibinfo {pages} {305} (\bibinfo {year} {1999})}\BibitemShut {NoStop}%
\bibitem [{\citenamefont {Hinohara}\ \emph {et~al.}(2013)\citenamefont {Hinohara}, \citenamefont {Kortelainen},\ and\ \citenamefont {Nazarewicz}}]{Hinohara2013}%
  \BibitemOpen
  \bibfield  {author} {\bibinfo {author} {\bibfnamefont {N.}~\bibnamefont {Hinohara}}, \bibinfo {author} {\bibfnamefont {M.}~\bibnamefont {Kortelainen}},\ and\ \bibinfo {author} {\bibfnamefont {W.}~\bibnamefont {Nazarewicz}},\ }\bibfield  {title} {\bibinfo {title} {Low-energy collective modes of deformed superfluid nuclei within the finite-amplitude method},\ }\href {https://doi.org/10.1103/PhysRevC.87.064309} {\bibfield  {journal} {\bibinfo  {journal} {Phys. Rev. C}\ }\textbf {\bibinfo {volume} {87}},\ \bibinfo {pages} {064309} (\bibinfo {year} {2013})}\BibitemShut {NoStop}%
\bibitem [{\citenamefont {Porro}\ \emph {et~al.}(2024)\citenamefont {Porro}, \citenamefont {Col\`o}, \citenamefont {Duguet}, \citenamefont {Gambacurta},\ and\ \citenamefont {Som\`a}}]{Porro2024}%
  \BibitemOpen
  \bibfield  {author} {\bibinfo {author} {\bibfnamefont {A.}~\bibnamefont {Porro}}, \bibinfo {author} {\bibfnamefont {G.}~\bibnamefont {Col\`o}}, \bibinfo {author} {\bibfnamefont {T.}~\bibnamefont {Duguet}}, \bibinfo {author} {\bibfnamefont {D.}~\bibnamefont {Gambacurta}},\ and\ \bibinfo {author} {\bibfnamefont {V.}~\bibnamefont {Som\`a}},\ }\bibfield  {title} {\bibinfo {title} {{Symmetry-restored Skyrme-random-phase-approximation calculations of the monopole strength in deformed nuclei}},\ }\href {https://doi.org/10.1103/PhysRevC.109.044315} {\bibfield  {journal} {\bibinfo  {journal} {Phys. Rev. C}\ }\textbf {\bibinfo {volume} {109}},\ \bibinfo {pages} {044315} (\bibinfo {year} {2024})}\BibitemShut {NoStop}%
\bibitem [{\citenamefont {Kanerva}\ and\ \citenamefont {Kortelainen}(2026)}]{Data_repository}%
  \BibitemOpen
  \bibfield  {author} {\bibinfo {author} {\bibfnamefont {M.}~\bibnamefont {Kanerva}}\ and\ \bibinfo {author} {\bibfnamefont {M.}~\bibnamefont {Kortelainen}},\ }\href {https://doi.org/10.17011/jyx/dataset/109498} {\bibinfo {title} {Transition strengths of octupole-deformed actinide nuclei}},\ \bibinfo {howpublished} {University of Jyväskylä} (\bibinfo {year} {2026})\BibitemShut {NoStop}%
\end{thebibliography}
\end{document}